\documentclass[letterpaper,10pt]{article} 

\usepackage{osameet3} 

\newcommand\authormark[1]{\textsuperscript{#1}}

\usepackage{amsmath,amssymb}
\usepackage{pgfplotstable}
\usepackage{subcaption}
\usepackage{siunitx}
\usepackage[colorlinks=true,bookmarks=false,citecolor=blue,urlcolor=blue]{hyperref} 
\definecolor{C1}{RGB}{031, 119, 180} 
\definecolor{C2}{RGB}{255, 127, 014} 
\definecolor{C3}{RGB}{044, 160, 044} 
\definecolor{C4}{RGB}{215, 039, 040} 
\definecolor{C6}{RGB}{148, 103, 189} 
\definecolor{C100}{RGB}{140, 086, 075} 
\definecolor{C7}{RGB}{227, 119, 194} 
\definecolor{C8}{RGB}{127, 127, 127} 
\definecolor{C9}{RGB}{188, 189, 034} 
\definecolor{C5}{RGB}{023, 190, 207} 

\definecolor{C11}{RGB}{174, 199, 232} 
\definecolor{C12}{RGB}{255, 187, 120} 
\definecolor{C13}{RGB}{152, 223, 138} 
\definecolor{C14}{RGB}{255, 152, 150} 
\definecolor{C15}{RGB}{197, 176, 213} 
\definecolor{C16}{RGB}{196, 156, 148} 
\definecolor{C17}{RGB}{247, 182, 210} 
\definecolor{C18}{RGB}{199, 199, 199} 
\definecolor{C19}{RGB}{219, 219, 141} 
\definecolor{C20}{RGB}{158, 218, 229} 
\usepackage{cite}
\usepackage[font=small]{caption}
\usetikzlibrary{backgrounds}
\usepackage{import}
\usepackage{tikz}
\usepackage{import}

\usetikzlibrary{positioning}
\usetikzlibrary{fit}
\usetikzlibrary{calc}
\usetikzlibrary{shapes}
\usetikzlibrary{math}
\usetikzlibrary{fpu}
\usetikzlibrary{decorations.markings}
\usetikzlibrary{decorations.pathreplacing}
\usetikzlibrary{arrows.meta}
\usetikzlibrary{intersections}

\makeatletter

\tikzset{outer sep=0}
\tikzset{inner sep=0}

\def\pgfaddtoshape#1#2{
	\begingroup
	\def\pgf@sm@shape@name{#1}%
	\let\anchor\pgf@sh@anchor
	#2%
	\endgroup
}

\newcommand{\anchorlet}[2]{
	\global\expandafter
	\let\csname pgf@anchor@\pgf@sm@shape@name @#1\expandafter\endcsname
	\csname pgf@anchor@\pgf@sm@shape@name @#2\endcsname
}

\pgfmathsetmacro{\NODESIZE}{42}
\pgfmathsetmacro{\NODETHICKNESS}{1.0}
\pgfmathsetmacro{\ROUNDEDCORNERS}{0.5mm}
\tikzset{node distance=0.5*\NODESIZE pt}

\def\FNODESIZE{\NODESIZE pt}

\tikzstyle{textstyle} = [text height=1.5ex, text depth=.5ex]
\tikzset{every label/.style=textstyle}

\tikzstyle{linestyle} = [line width = \NODETHICKNESS, rounded corners = \ROUNDEDCORNERS]
\tikzstyle{arrowstyle} = [>=stealth, linestyle]
\tikzstyle{<--} = [<-, arrowstyle]
\tikzstyle{-->} = [->, arrowstyle]
\tikzstyle{->-} = [linestyle, decoration={markings,	mark=at position 0.5 with {\arrow[arrowstyle]{>}}}, postaction={decorate}] 
\tikzstyle{-<-} = [linestyle, decoration={markings,	mark=at position 0.5 with {\arrow[arrowstyle]{<}}}, postaction={decorate}] 

\tikzset{   
    -A-/.style args={#1}{%
        linestyle, 
        decoration={markings, mark=at position #1 with {\arrow[>=Triangle, scale=.025*\NODESIZE]{>}}},
        postaction={decorate}
    },
    -A-/.default = {0.75}
}

\tikzset{   
    -AA-/.style args={#1, #2, #3}{%
        linestyle,
        decoration={markings, mark=at position #1 with {
            \node[amp={color=#2, width=#3, height=#3}, rotate=\pgfdecoratedangle] at (0,0) (inline_amp) {};
        }},
        postaction={decorate}
    },
    -AA-/.default = {0.5, E, 0.15}
}

\tikzstyle{<-->} = [<->, arrowstyle]
\tikzstyle{---} = [arrowstyle]

\tikzset{
    -PC-/.style args={#1}{
        linestyle, 
        decoration={markings, mark=at position #1 with {
            \draw[---,FO]
                (.05*\NODESIZE*\pgflinewidth,.05*\NODESIZE*\pgflinewidth) circle[radius=.05*\NODESIZE*\pgflinewidth];
            \draw[---,FO]
                (-.05*\NODESIZE*\pgflinewidth,.05*\NODESIZE*\pgflinewidth) circle[radius=.05*\NODESIZE*\pgflinewidth];
            \draw[---,FO]
                (0,-.05*\NODESIZE*\pgflinewidth) circle[radius=.05*\NODESIZE*\pgflinewidth];}},
        postaction={decorate}
        },
        -PC-/.default = {0.5}
}

\definecolor{C0}{RGB}{031, 119, 180} 
\definecolor{C1}{RGB}{031, 119, 180} 
\definecolor{C2}{RGB}{255, 127, 014} 
\definecolor{C3}{RGB}{044, 160, 044} 
\definecolor{C4}{RGB}{215, 039, 040} 
\definecolor{C6}{RGB}{148, 103, 189} 
\definecolor{C100}{RGB}{140, 086, 075} 
\definecolor{C7}{RGB}{227, 119, 194} 
\definecolor{C8}{RGB}{127, 127, 127} 
\definecolor{C9}{RGB}{188, 189, 034} 
\definecolor{C5}{RGB}{023, 190, 207} 

\definecolor{C0l}{RGB}{174, 199, 232} 
\definecolor{C11}{RGB}{174, 199, 232} 
\definecolor{C12}{RGB}{255, 187, 120} 
\definecolor{C13}{RGB}{152, 223, 138} 
\definecolor{C14}{RGB}{255, 152, 150} 
\definecolor{C15}{RGB}{197, 176, 213} 
\definecolor{C16}{RGB}{196, 156, 148} 
\definecolor{C17}{RGB}{247, 182, 210} 
\definecolor{C18}{RGB}{199, 199, 199} 
\definecolor{C19}{RGB}{219, 219, 141} 
\definecolor{C20}{RGB}{158, 218, 229} 

\definecolor{Snow}{HTML}{FBFBFB} 			
\definecolor{TUeRed}{RGB}{200, 25, 25}		
\definecolor{TUeGreen}{RGB}{25, 200, 113}	
\definecolor{TUeBlue}{RGB}{25, 113, 200}	

\definecolor{O}{RGB}{031, 119, 180} 	
\definecolor{Ol}{RGB}{174, 199, 232} 	
\definecolor{E}{RGB}{255, 127, 014} 	
\definecolor{El}{RGB}{255, 187, 120} 	
\definecolor{D}{RGB}{148, 103, 189}     
\definecolor{Dl}{RGB}{197, 176, 213} 	
\definecolor{EO}{RGB}{215, 039, 040} 	
\definecolor{EOl}{RGB}{255, 152, 150}   

\tikzstyle{FW} = [fill=white]			
\tikzstyle{FB} = [fill=white]			
\tikzstyle{FO} = [fill=C0l, draw=C0]	
\tikzstyle{FE} = [fill=C1l, draw=C1]	
\tikzstyle{FD} = [fill=C2l, draw=C2]	

\def\direce{e}
\def\direcw{w}
\def\direcn{n}
\def\direcs{s}

\def\flipfalse{0}
\newcount\portcount
\pgfdeclareshape{basic}{
	\inheritsavedanchors[from=rectangle] 
	
	\inheritanchor[from=rectangle]{center}
	\inheritanchor[from=rectangle]{mid}		
	\inheritanchor[from=rectangle]{base}	
	\inheritanchor[from=rectangle]{north}
	\inheritanchor[from=rectangle]{south}		
	\inheritanchor[from=rectangle]{west}		
	\inheritanchor[from=rectangle]{mid west}				
	\inheritanchor[from=rectangle]{base west}		
	\inheritanchor[from=rectangle]{north west}		
	\inheritanchor[from=rectangle]{south west}		
	\inheritanchor[from=rectangle]{east}
	\inheritanchor[from=rectangle]{mid east}
	\inheritanchor[from=rectangle]{base east}	
	\inheritanchor[from=rectangle]{north east}
	\inheritanchor[from=rectangle]{south east}	
	
	\inheritanchorborder[from=rectangle]	

	\inheritbackgroundpath[from=rectangle]	
}

\pgfaddtoshape{basic}{
	\anchor{west south west}{
		\pgf@process{\northeast}
		\pgf@ya=.5\pgf@y
		\pgf@process{\southwest}
		\pgf@y=1.5\pgf@y
		\advance\pgf@y by \pgf@ya
		\pgf@y=.5\pgf@y
	}
	\anchor{west north west}{
		\pgf@process{\northeast}
		\pgf@ya=1.5\pgf@y
		\pgf@process{\southwest}
		\pgf@y=.5\pgf@y
		\advance\pgf@y by \pgf@ya
		\pgf@y=.5\pgf@y
	}
	\anchor{east north east}{
		\pgf@process{\southwest}
		\pgf@ya=.5\pgf@y
		\pgf@process{\northeast}
		\pgf@y=1.5\pgf@y
		\advance\pgf@y by \pgf@ya
		\pgf@y=.5\pgf@y
	}
	\anchor{east south east}{
		\pgf@process{\southwest}
		\pgf@ya=1.5\pgf@y
		\pgf@process{\northeast}
		\pgf@y=.5\pgf@y
		\advance\pgf@y by \pgf@ya
		\pgf@y=.5\pgf@y
	}
	\anchor{north north west}{
		\pgf@process{\southwest}
		\pgf@xa=1.5\pgf@x
		\pgf@process{\northeast}
		\pgf@x=.5\pgf@x
		\advance\pgf@x by \pgf@xa
		\pgf@x=.5\pgf@x
	}
	\anchor{north north east}{
		\pgf@process{\southwest}
		\pgf@xa=.5\pgf@x
		\pgf@process{\northeast}
		\pgf@x=1.5\pgf@x
		\advance\pgf@x by \pgf@xa
		\pgf@x=.5\pgf@x
	}
	\anchor{south south west}{
		\pgf@process{\northeast}
		\pgf@xa=.5\pgf@x
		\pgf@process{\southwest}
		\pgf@x=1.5\pgf@x
		\advance\pgf@x by \pgf@xa
		\pgf@x=.5\pgf@x
	}
	\anchor{south south east}{
		\pgf@process{\northeast}
		\pgf@xa=1.5\pgf@x
		\pgf@process{\southwest}
		\pgf@x=.5\pgf@x
		\advance\pgf@x by \pgf@xa
		\pgf@x=.5\pgf@x
	}
}

\pgfaddtoshape{basic}{
	\anchorlet{c}{center}		
	\anchorlet{n}{north}
	\anchorlet{e}{east}
	\anchorlet{s}{south}
	\anchorlet{w}{west}			
	\anchorlet{se}{south east}
	\anchorlet{sw}{south west}
	\anchorlet{ne}{north east}
	\anchorlet{nw}{north west}
	\anchorlet{wsw}{west south west}
	\anchorlet{wnw}{west north west}
	\anchorlet{ene}{east north east}
	\anchorlet{ese}{east south east}
	\anchorlet{nnw}{north north west}
	\anchorlet{nne}{north north east}
	\anchorlet{ssw}{south south west}
	\anchorlet{sse}{south south east}
}	

\pgfdeclareshape{ampshape}{
	\savedmacro\direction{
		\edef\direction{\pgfkeysvalueof{/tikz/ampkeys/direction}}%
	}
	\saveddimen\minwidth{
		\pgfmathsetlength\pgf@x{\pgfshapeminwidth}%
	}
	\saveddimen\minheight{
		\pgfmathsetlength\pgf@x{\pgfshapeminheight}%
	}
	\inheritsavedanchors[from=basic] 
	
	\inheritanchor[from=basic]{center}
	\inheritanchor[from=basic]{mid}		
	\inheritanchor[from=basic]{base}	
	\inheritanchor[from=basic]{north}
	\inheritanchor[from=basic]{south}		
	\inheritanchor[from=basic]{west}		
	\inheritanchor[from=basic]{mid west}				
	\inheritanchor[from=basic]{base west}		
	\inheritanchor[from=basic]{north west}		
	\inheritanchor[from=basic]{south west}		
	\inheritanchor[from=basic]{east}
	\inheritanchor[from=basic]{mid east}
	\inheritanchor[from=basic]{base east}	
	\inheritanchor[from=basic]{north east}
	\inheritanchor[from=basic]{south east}
	\inheritanchor[from=basic]{west south west}%
	\inheritanchor[from=basic]{west north west}%
	\inheritanchor[from=basic]{east north east}%
	\inheritanchor[from=basic]{east south east}%
	\inheritanchor[from=basic]{north north west}%
	\inheritanchor[from=basic]{north north east}%
	\inheritanchor[from=basic]{south south west}%
	\inheritanchor[from=basic]{south south east}%
	
	\inheritanchorborder[from=basic]	
	\inheritbackgroundpath[from=basic]

    \pgfutil@g@addto@macro\pgf@sh@s@ampshape{%
        \pgfutil@ifundefined{pgf@anchor@ampshape@in0}{
	        \expandafter\xdef\csname pgf@anchor@ampshape@in0\endcsname{%
	            \noexpand\ampshape@port{0}
	        }%
	    }{}%
        \pgfutil@ifundefined{pgf@anchor@ampshape@in}{
	        \expandafter\xdef\csname pgf@anchor@ampshape@in\endcsname{%
	            \noexpand\ampshape@port{0}
	        }%
	    }{}%
        \pgfutil@ifundefined{pgf@anchor@ampshape@out0}{
	        \expandafter\xdef\csname pgf@anchor@ampshape@out0\endcsname{%
	            \noexpand\ampshape@port{1}
	        }%
	    }{}%
        \pgfutil@ifundefined{pgf@anchor@ampshape@out}{
	        \expandafter\xdef\csname pgf@anchor@ampshape@out\endcsname{%
	            \noexpand\ampshape@port{1}
	        }%
	    }{}%
	}
}

\def\ampshape@port#1{
    \northeast	

    \ifnum#1=0	
	    \if\direction\direce
			\pgf@x=-\pgf@x
		    \pgf@ya= \pgf@y
		    \pgfmathsetlength{\pgf@y}{\pgf@ya-0.5*\minheight}%
		\fi
	    \if\direction\direcw
			\pgf@x=\pgf@x
		    \pgf@ya= \pgf@y
		    \pgfmathsetlength{\pgf@y}{\pgf@ya-0.5*\minheight}%
		\fi
	    \if\direction\direcn
			\pgf@y=-\pgf@y
		    \pgf@xa=\pgf@x
		    \pgfmathsetlength{\pgf@x}{\pgf@xa-0.5*\minwidth}%
		\fi
	    \if\direction\direcs
			\pgf@y=\pgf@y
		    \pgf@xa= \pgf@x
		    \pgfmathsetlength{\pgf@x}{\pgf@xa-0.5*\minwidth}%
		\fi
	\else	
	    \if\direction\direce
			\pgf@x=\pgf@x
		    \pgf@ya= \pgf@y
		    \pgfmathsetlength{\pgf@y}{\pgf@ya-0.5*\minheight}%
		\fi
	    \if\direction\direcw
			\pgf@x=-\pgf@x
		    \pgf@ya= \pgf@y
		    \pgfmathsetlength{\pgf@y}{\pgf@ya-0.5*\minheight}%
		\fi
	    \if\direction\direcn
			\pgf@y=\pgf@y
		    \pgf@xa= \pgf@x
		    \pgfmathsetlength{\pgf@x}{\pgf@xa-0.5*\minwidth}%
		\fi
	    \if\direction\direcs
			\pgf@y=-\pgf@y
		    \pgf@xa= \pgf@x
		    \pgfmathsetlength{\pgf@x}{\pgf@xa-0.5*\minwidth}%
		\fi
	\fi
}

\pgfaddtoshape{ampshape}{
	\anchorlet{c}{center}		
	\anchorlet{n}{north}
	\anchorlet{e}{east}
	\anchorlet{s}{south}
	\anchorlet{w}{west}			
	\anchorlet{se}{south east}
	\anchorlet{sw}{south west}
	\anchorlet{ne}{north east}
	\anchorlet{nw}{north west}
	\anchorlet{wsw}{west south west}
	\anchorlet{wnw}{west north west}
	\anchorlet{ene}{east north east}
	\anchorlet{ese}{east south east}
	\anchorlet{nnw}{north north west}
	\anchorlet{nne}{north north east}
	\anchorlet{ssw}{south south west}
	\anchorlet{sse}{south south east}
}

\tikzset{
	/tikz/ampkeys/.cd,
	height/.initial=0.5,
	width/.initial=0.5,
	color/.initial=O,
	direction/.initial=e,
	linestyle/.initial={linestyle, rounded corners = 0},
	/tikz/amp/.code={
		\pgfqkeys{/tikz/ampkeys}{#1}%
		\tikzset{/tikz/ampkeys/drawer/.expanded=%
			{\pgfkeysvalueof{/tikz/ampkeys/direction}}%
			{\pgfkeysvalueof{/tikz/ampkeys/height}}%
			{\pgfkeysvalueof{/tikz/ampkeys/width}}%
			{\pgfkeysvalueof{/tikz/ampkeys/color}}%
			{\pgfkeysvalueof{/tikz/ampkeys/linestyle}}%
		}
	},
	/tikz/ampkeys/drawer/.code n args={5}{%
		\tikzset{
			ampshape,
			minimum height=#2*\NODESIZE,
			minimum width=#3*\NODESIZE,
			append after command={
				\pgfextra{\let\bdr=\tikzlastnode%
				\if#1e
					\draw[draw=#4, fill=#4l, #5] (\bdr.sw) to (\bdr.nw) to (\bdr.e) to cycle {};
				\fi
				\if#1w
					\draw[draw=#4, fill=#4l, #5] (\bdr.se) to (\bdr.ne) to (\bdr.w) to cycle {};
				\fi
				\if#1n
					\draw[draw=#4, fill=#4l, #5] (\bdr.se) to (\bdr.sw) to (\bdr.n) to cycle {};
				\fi
				\if#1s
					\draw[draw=#4, fill=#4l, #5] (\bdr.ne) to (\bdr.nw) to (\bdr.s) to cycle {};
				\fi
				}
			}
		}
	},
}
\newcount\portcount

\pgfdeclareshape{aomshape}{
	\savedmacro\direction{
		\edef\direction{\pgfkeysvalueof{/tikz/aomkeys/direction}}%
	}
	\saveddimen\minwidth{
		\pgfmathsetlength\pgf@x{\pgfshapeminwidth}%
	}
	\saveddimen\minheight{
		\pgfmathsetlength\pgf@x{\pgfshapeminheight}%
	}
	\inheritsavedanchors[from=basic]

	\inheritanchor[from=basic]{center}
	\inheritanchor[from=basic]{mid}
	\inheritanchor[from=basic]{base}
	\inheritanchor[from=basic]{north}
	\inheritanchor[from=basic]{south}
	\inheritanchor[from=basic]{west}
	\inheritanchor[from=basic]{mid west}
	\inheritanchor[from=basic]{base west}
	\inheritanchor[from=basic]{north west}
	\inheritanchor[from=basic]{south west}
	\inheritanchor[from=basic]{east}
	\inheritanchor[from=basic]{mid east}
	\inheritanchor[from=basic]{base east}
	\inheritanchor[from=basic]{north east}
	\inheritanchor[from=basic]{south east}
	\inheritanchor[from=basic]{west south west}%
	\inheritanchor[from=basic]{west north west}%
	\inheritanchor[from=basic]{east north east}%
	\inheritanchor[from=basic]{east south east}%
	\inheritanchor[from=basic]{north north west}%
	\inheritanchor[from=basic]{north north east}%
	\inheritanchor[from=basic]{south south west}%
	\inheritanchor[from=basic]{south south east}%

	\inheritanchorborder[from=basic]
	\inheritbackgroundpath[from=basic]

	\pgfutil@g@addto@macro\pgf@sh@s@aomshape{%
		\pgfutil@ifundefined{pgf@anchor@aomshape@in0}{
			\expandafter\xdef\csname pgf@anchor@aomshape@in0\endcsname{%
				\noexpand\aomshape@port{0}
			}%
		}{}%
		\pgfutil@ifundefined{pgf@anchor@aomshape@in}{
			\expandafter\xdef\csname pgf@anchor@aomshape@in\endcsname{%
				\noexpand\aomshape@port{0}
			}%
		}{}%
		\pgfutil@ifundefined{pgf@anchor@aomshape@out0}{
			\expandafter\xdef\csname pgf@anchor@aomshape@out0\endcsname{%
				\noexpand\aomshape@port{1}
			}%
		}{}%
		\pgfutil@ifundefined{pgf@anchor@aomshape@out}{
			\expandafter\xdef\csname pgf@anchor@aomshape@out\endcsname{%
				\noexpand\aomshape@port{1}
			}%
		}{}%
	}
}

\def\aomshape@port#1{
	\northeast	

	\ifnum#1=0	
		\if\direction\direce
			\pgf@x=-\pgf@x
			\pgf@ya= \pgf@y
			\pgfmathsetlength{\pgf@y}{\pgf@ya-0.5*\minheight}%
		\fi
		\if\direction\direcw
			\pgf@x=\pgf@x
			\pgf@ya= \pgf@y
			\pgfmathsetlength{\pgf@y}{\pgf@ya-0.5*\minheight}%
		\fi
		\if\direction\direcn
			\pgf@y=-\pgf@y
			\pgf@xa=\pgf@x
			\pgfmathsetlength{\pgf@x}{\pgf@xa-0.5*\minwidth}%
		\fi
		\if\direction\direcs
			\pgf@y=\pgf@y
			\pgf@xa= \pgf@x
			\pgfmathsetlength{\pgf@x}{\pgf@xa-0.5*\minwidth}%
		\fi
	\else	
		\if\direction\direce
			\pgf@x=\pgf@x
			\pgf@ya= \pgf@y
			\pgfmathsetlength{\pgf@y}{\pgf@ya-0.5*\minheight}%
		\fi
		\if\direction\direcw
			\pgf@x=-\pgf@x
			\pgf@ya= \pgf@y
			\pgfmathsetlength{\pgf@y}{\pgf@ya-0.5*\minheight}%
		\fi
		\if\direction\direcn
			\pgf@y=\pgf@y
			\pgf@xa= \pgf@x
			\pgfmathsetlength{\pgf@x}{\pgf@xa-0.5*\minwidth}%
		\fi
		\if\direction\direcs
			\pgf@y=-\pgf@y
			\pgf@xa= \pgf@x
			\pgfmathsetlength{\pgf@x}{\pgf@xa-0.5*\minwidth}%
		\fi
	\fi
}

\pgfaddtoshape{aomshape}{
	\anchorlet{c}{center}
	\anchorlet{n}{north}
	\anchorlet{e}{east}
	\anchorlet{s}{south}
	\anchorlet{w}{west}
	\anchorlet{se}{south east}
	\anchorlet{sw}{south west}
	\anchorlet{ne}{north east}
	\anchorlet{nw}{north west}
	\anchorlet{wsw}{west south west}
	\anchorlet{wnw}{west north west}
	\anchorlet{ene}{east north east}
	\anchorlet{ese}{east south east}
	\anchorlet{nnw}{north north west}
	\anchorlet{nne}{north north east}
	\anchorlet{ssw}{south south west}
	\anchorlet{sse}{south south east}
}

\tikzset{
/tikz/aomkeys/.cd,
size/.initial=1,
circlesize/.initial=1,
color/.initial=O,
direction/.initial=e,
linestyle/.initial={linestyle, inner sep=0.5mm},
fillgradient/.initial=O,
/tikz/aom/.code={
\pgfqkeys{/tikz/aomkeys}{#1}%
\tikzset{/tikz/aomkeys/drawer/.expanded=%
	{\pgfkeysvalueof{/tikz/aomkeys/size}}%
	{\pgfkeysvalueof{/tikz/aomkeys/color}}%
	{\pgfkeysvalueof{/tikz/aomkeys/linestyle}}%
\if\pgfkeysvalueof{/tikz/aomkeys/direction}e
	{0}%
\fi
\if\pgfkeysvalueof{/tikz/aomkeys/direction}w
	{0}%
\fi
\if\pgfkeysvalueof{/tikz/aomkeys/direction}n
	{1}%
\fi
\if\pgfkeysvalueof{/tikz/aomkeys/direction}s
	{1}%
\fi
{\pgfkeysvalueof{/tikz/aomkeys/direction}}%
{\pgfkeysvalueof{/tikz/aomkeys/circlesize}}%
{\pgfkeysvalueof{/tikz/aomkeys/fillgradient}}%
}
},
/tikz/aomkeys/drawer/.code n args={7}{%
		\tikzset{
			aomshape,
			draw,
			minimum height = #1*\NODESIZE,
			minimum width = #1*\NODESIZE,
			#2,
			#3,
			append after command={
					\pgfextra{\let\bdr=\tikzlastnode%
						\node[#7, fit=(\bdr.nw)(\bdr.se)] (boxgradient){};

						\node[coordinate] at ($(\bdr.in)!0.25!(\bdr.out)$) (circlein){};
						\node[coordinate] at ($(\bdr.in)!0.75!(\bdr.out)$) (circleout){};

						\ifnum#4>0
							\node[coordinate] at (circleout -| \bdr.nne) (circleouttop){};
						\else
							\node[coordinate] at (circleout |- \bdr.ene) (circleouttop){};
						\fi

						\draw[---, #2, #3, fill] (\bdr.in) to (circlein) circle (0.05*#6);
						\draw[---, #2, #3, fill] (\bdr.out) to (circleout) circle (0.05*#6);

						\draw[---, #2, #3] (circlein) to (circleouttop){};

					}
				}
		}
	},
}
\newcount\portcount

\pgfdeclareshape{boxshape}{
	\savedmacro\nin{
		\edef\nin{\pgfkeysvalueof{/tikz/boxkeys/nin}}%
	}
	\savedmacro\nout{
		\edef\nout{\pgfkeysvalueof{/tikz/boxkeys/nout}}%
	}
	\savedmacro\direction{
		\edef\direction{\pgfkeysvalueof{/tikz/boxkeys/direction}}%
	}
	\inheritsavedanchors[from=basic]

	\inheritanchor[from=basic]{center}
	\inheritanchor[from=basic]{mid}
	\inheritanchor[from=basic]{base}
	\inheritanchor[from=basic]{north}
	\inheritanchor[from=basic]{south}
	\inheritanchor[from=basic]{west}
	\inheritanchor[from=basic]{mid west}
	\inheritanchor[from=basic]{base west}
	\inheritanchor[from=basic]{north west}
	\inheritanchor[from=basic]{south west}
	\inheritanchor[from=basic]{east}
	\inheritanchor[from=basic]{mid east}
	\inheritanchor[from=basic]{base east}
	\inheritanchor[from=basic]{north east}
	\inheritanchor[from=basic]{south east}
	\inheritanchor[from=basic]{west south west}%
	\inheritanchor[from=basic]{west north west}%
	\inheritanchor[from=basic]{east north east}%
	\inheritanchor[from=basic]{east south east}%
	\inheritanchor[from=basic]{north north west}%
	\inheritanchor[from=basic]{north north east}%
	\inheritanchor[from=basic]{south south west}%
	\inheritanchor[from=basic]{south south east}%

	\inheritanchorborder[from=basic]
	\inheritbackgroundpath[from=basic]

	\pgfutil@g@addto@macro\pgf@sh@s@boxshape{%
		\pgfmathsetcount{\portcount}{0}
		\pgfmathloop%
		\ifnum\the\portcount<\nin
		\pgfutil@ifundefined{pgf@anchor@boxshape@in\the\portcount}{
			\expandafter\xdef\csname pgf@anchor@boxshape@in\the\portcount\endcsname{%
				\noexpand\boxshape@port[\the\portcount]{0}
			}%
		}{}%
		\ifnum\the\portcount=0
			\pgfutil@ifundefined{pgf@anchor@boxshape@in}{%
				\expandafter\xdef\csname pgf@anchor@boxshape@in\endcsname{%
					\noexpand\boxshape@port[\the\portcount]{0}
				}%
			}{}%
		\fi
		\pgfmathaddtocount{\portcount}{1}	
		\repeatpgfmathloop
		\pgfmathsetcount{\portcount}{0}
		\pgfmathloop%
		\ifnum\the\portcount<\nout
		\pgfutil@ifundefined{pgf@anchor@boxshape@out\the\portcount}{%
			\expandafter\xdef\csname pgf@anchor@boxshape@out\the\portcount\endcsname{%
				\noexpand\boxshape@port[\the\portcount]{1}
			}%
		}{}%
		\ifnum\the\portcount=0
			\pgfutil@ifundefined{pgf@anchor@boxshape@out}{%
				\expandafter\xdef\csname pgf@anchor@boxshape@out\endcsname{%
					\noexpand\boxshape@port[\the\portcount]{1}
				}%
			}{}%
		\fi
		\pgfmathaddtocount{\portcount}{1}	
		\repeatpgfmathloop
	}
}

\def\boxshape@port[#1]#2{
	\northeast \pgf@xa=\pgf@x \pgf@ya=\pgf@y
	\southwest \pgf@xb=\pgf@x \pgf@yb=\pgf@y

	\ifnum#2=0	
		\if\direction\direce	
			\pgf@x=\pgf@xb
			\pgf@yc=\pgf@ya \advance\pgf@yc by -\pgf@yb	
			\pgfmathsetlength{\pgf@y}{\pgf@ya-(#1 + 0.5)*(\pgf@yc/\nin)}%
		\fi
		\if\direction\direcw
			\pgf@x=\pgf@xa
			\pgf@yc=\pgf@ya \advance\pgf@yc by -\pgf@yb	
			\pgfmathsetlength{\pgf@y}{\pgf@ya-(#1 + 0.5)*(\pgf@yc/\nin)}%
		\fi
		\if\direction\direcn
			\pgf@y=\pgf@yb
			\pgf@xc=\pgf@xa \advance\pgf@xc by -\pgf@xb	
			\pgfmathsetlength{\pgf@x}{\pgf@xb+(#1 + 0.5)*(\pgf@xc/\nin)}%
		\fi
		\if\direction\direcs
			\pgf@y=\pgf@ya
			\pgf@xc=\pgf@xa \advance\pgf@xc by -\pgf@xb	
			\pgfmathsetlength{\pgf@x}{\pgf@xb+(#1 + 0.5)*(\pgf@xc/\nin)}%
		\fi
	\else	
		\if\direction\direce	
			\pgf@x=\pgf@xa
			\pgf@yc=\pgf@ya \advance\pgf@yc by -\pgf@yb	
			\pgfmathsetlength{\pgf@y}{\pgf@ya-(#1 + 0.5)*(\pgf@yc/\nout)}%
		\fi
		\if\direction\direcw
			\pgf@x=\pgf@xb
			\pgf@yc=\pgf@ya \advance\pgf@yc by -\pgf@yb	
			\pgfmathsetlength{\pgf@y}{\pgf@ya-(#1 + 0.5)*(\pgf@yc/\nout)}%
		\fi
		\if\direction\direcn
			\pgf@y=\pgf@ya
			\pgf@xc=\pgf@xa \advance\pgf@xc by -\pgf@xb	
			\pgfmathsetlength{\pgf@x}{\pgf@xb+(#1 + 0.5)*(\pgf@xc/\nout)}%
		\fi
		\if\direction\direcs
			\pgf@y=\pgf@yb
			\pgf@xc=\pgf@xa \advance\pgf@xc by -\pgf@xb	
			\pgfmathsetlength{\pgf@x}{\pgf@xb+(#1 + 0.5)*(\pgf@xc/\nout)}%
		\fi
	\fi
}

\pgfaddtoshape{boxshape}{
	\anchorlet{c}{center}
	\anchorlet{n}{north}
	\anchorlet{e}{east}
	\anchorlet{s}{south}
	\anchorlet{w}{west}
	\anchorlet{se}{south east}
	\anchorlet{sw}{south west}
	\anchorlet{ne}{north east}
	\anchorlet{nw}{north west}
	\anchorlet{wsw}{west south west}
	\anchorlet{wnw}{west north west}
	\anchorlet{ene}{east north east}
	\anchorlet{ese}{east south east}
	\anchorlet{nnw}{north north west}
	\anchorlet{nne}{north north east}
	\anchorlet{ssw}{south south west}
	\anchorlet{sse}{south south east}
}

\tikzset{
/tikz/boxkeys/.cd,
height/.initial=0.5,
width/.initial=1,
color/.initial=O,
direction/.initial=e,
linestyle/.initial={linestyle, inner sep=0.5mm},
nin/.initial=1,
nout/.initial=1,
draw/.initial=1,
/tikz/box/.code={
\pgfqkeys{/tikz/boxkeys}{#1}%
\tikzset{/tikz/boxkeys/drawer/.expanded=%
\if\pgfkeysvalueof{/tikz/boxkeys/direction}e
	{\pgfkeysvalueof{/tikz/boxkeys/width}}%
	{\pgfkeysvalueof{/tikz/boxkeys/height}}%
	{0}%
	{-90}%
\fi
\if\pgfkeysvalueof{/tikz/boxkeys/direction}w
	{\pgfkeysvalueof{/tikz/boxkeys/width}}%
	{\pgfkeysvalueof{/tikz/boxkeys/height}}%
	{0}%
	{90}%
\fi
\if\pgfkeysvalueof{/tikz/boxkeys/direction}n
	{\pgfkeysvalueof{/tikz/boxkeys/width}}%
	{\pgfkeysvalueof{/tikz/boxkeys/height}}%
	{1}%
	{0}%
\fi
\if\pgfkeysvalueof{/tikz/boxkeys/direction}s
	{\pgfkeysvalueof{/tikz/boxkeys/width}}%
	{\pgfkeysvalueof{/tikz/boxkeys/height}}%
	{1}%
	{180}%
\fi
{\pgfkeysvalueof{/tikz/boxkeys/color}}%
{\pgfkeysvalueof{/tikz/boxkeys/linestyle}}%
\ifnum\pgfkeysvalueof{/tikz/boxkeys/draw}>0%
	{draw}%
\else
	{}
\fi
}
},
/tikz/boxkeys/drawer/.code n args={7}{%
		\tikzset{
			boxshape,
			#7,
			#6,
			#5,
			minimum height=
			\ifnum#3>0	
				#1*\NODESIZE
			\else
				#2*\NODESIZE
			\fi
			,minimum width=
			\ifnum#3>0
				#2*\NODESIZE
			\else
				#1*\NODESIZE
			\fi
		}
	},
}
\pgfdeclareshape{beshape}{ 
    \savedmacro\nports{
        \edef\nports{\pgfkeysvalueof{/tikz/bekeys/nports}}%
    }
    \savedmacro\direction{
        \edef\direction{\pgfkeysvalueof{/tikz/bekeys/direction}}%
    }
    \savedmacro\inverted{
        \edef\inverted{\pgfkeysvalueof{/tikz/bekeys/inverted}}%
    }
    \savedmacro\ninports{
        \ifnum\inverted=0
        \edef\ninports{\nports}
        \else
        \edef\ninports{1}%
        \fi
    }
    \savedmacro\noutports{
        \ifnum\inverted=0
        \edef\noutports{1}%
        \else
        \edef\noutports{\nports}%
        \fi
    }
    \inheritsavedanchors[from=basic]

    \inheritanchor[from=basic]{center}
    \inheritanchor[from=basic]{mid}
    \inheritanchor[from=basic]{base}
    \inheritanchor[from=basic]{north}
    \inheritanchor[from=basic]{south}
    \inheritanchor[from=basic]{west}
    \inheritanchor[from=basic]{mid west}
    \inheritanchor[from=basic]{base west}
    \inheritanchor[from=basic]{north west}
    \inheritanchor[from=basic]{south west}
    \inheritanchor[from=basic]{east}
    \inheritanchor[from=basic]{mid east}
    \inheritanchor[from=basic]{base east}
    \inheritanchor[from=basic]{north east}
    \inheritanchor[from=basic]{south east}
    \inheritanchor[from=basic]{west south west}%
    \inheritanchor[from=basic]{west north west}%
    \inheritanchor[from=basic]{east north east}%
    \inheritanchor[from=basic]{east south east}%
    \inheritanchor[from=basic]{north north west}%
    \inheritanchor[from=basic]{north north east}%
    \inheritanchor[from=basic]{south south west}%
    \inheritanchor[from=basic]{south south east}%

    \inheritanchorborder[from=basic]
    \inheritbackgroundpath[from=basic]

    \pgfutil@g@addto@macro\pgf@sh@s@beshape{%
        \pgfmathsetcount{\portcount}{0}
        \pgfmathloop%
        \ifnum\the\portcount<\nports
        \ifnum\the\portcount<\ninports
        \pgfutil@ifundefined{pgf@anchor@beshape@in\the\portcount}{
            \expandafter\xdef\csname pgf@anchor@beshape@in\the\portcount\endcsname{%
                \noexpand\beshape@port[\the\portcount]{0}
            }%
        }{}%
        \ifnum\the\portcount=0
        \pgfutil@ifundefined{pgf@anchor@beshape@in}{%
            \expandafter\xdef\csname pgf@anchor@beshape@in\endcsname{%
                \noexpand\beshape@port[\the\portcount]{0}
            }%
        }{}%
        \fi
        \fi
        \ifnum\the\portcount<\noutports
        \pgfutil@ifundefined{pgf@anchor@beshape@out\the\portcount}{%
            \expandafter\xdef\csname pgf@anchor@beshape@out\the\portcount\endcsname{%
                \noexpand\beshape@port[\the\portcount]{1}
            }%
        }{}%
        \ifnum\the\portcount=0
        \pgfutil@ifundefined{pgf@anchor@beshape@out}{%
            \expandafter\xdef\csname pgf@anchor@beshape@out\endcsname{%
                \noexpand\beshape@port[\the\portcount]{1}
            }%
        }{}%
        \fi
        \fi
        \pgfmathaddtocount{\portcount}{1}    
        \repeatpgfmathloop
    }
}

\def\beshape@port[#1]#2{
    \northeast \pgf@xa=\pgf@x \pgf@ya=\pgf@y
    \southwest \pgf@xb=\pgf@x \pgf@yb=\pgf@y

    \ifnum#2=0
    \ifnum\inverted=0
    \def\chooseports{0}
    \else
    \def\chooseports{1}
    \fi
    \else
    \ifnum\inverted=0
    \def\chooseports{1}
    \else
    \def\chooseports{0}
    \fi
    \fi

    \ifnum\chooseports=0    
    \if\direction\direce
    \pgf@x=\pgf@xb
    \pgf@yc=\pgf@ya \advance\pgf@yc by -\pgf@yb    
    \pgfmathsetlength{\pgf@y}{\pgf@ya-(#1 + 0.5)*(\pgf@yc/\nports)}%
    \fi
    \if\direction\direcw
    \pgf@x=\pgf@xa
    \pgf@yc=\pgf@ya \advance\pgf@yc by -\pgf@yb    
    \pgfmathsetlength{\pgf@y}{\pgf@ya-(#1 + 0.5)*(\pgf@yc/\nports)}%
    \fi
    \if\direction\direcn
    \pgf@y=\pgf@yb
    \pgf@xc=\pgf@xa \advance\pgf@xc by -\pgf@xb    
    \pgfmathsetlength{\pgf@x}{\pgf@xb+(#1 + 0.5)*(\pgf@xc/\nports)}%
    \fi
    \if\direction\direcs
    \pgf@y=\pgf@ya
    \pgf@xc=\pgf@xa \advance\pgf@xc by -\pgf@xb    
    \pgfmathsetlength{\pgf@x}{\pgf@xb+(#1 + 0.5)*(\pgf@xc/\nports)}%
    \fi
    \else    
    \if\direction\direce
    \pgf@x=\pgf@xa
    \pgf@yc=\pgf@ya \advance\pgf@yc by -\pgf@yb    
    \pgfmathsetlength{\pgf@y}{\pgf@ya-0.5\pgf@yc}%
    \fi
    \if\direction\direcw
    \pgf@x=\pgf@xb
    \pgf@yc=\pgf@ya \advance\pgf@yc by -\pgf@yb    
    \pgfmathsetlength{\pgf@y}{\pgf@ya-0.5\pgf@yc}%
    \fi
    \if\direction\direcn
    \pgf@y=\pgf@ya
    \pgf@xc=\pgf@xa \advance\pgf@xc by -\pgf@xb    
    \pgfmathsetlength{\pgf@x}{\pgf@xa-0.5\pgf@xc}%
    \fi
    \if\direction\direcs
    \pgf@y=\pgf@yb
    \pgf@xc=\pgf@xa \advance\pgf@xc by -\pgf@xb    
    \pgfmathsetlength{\pgf@x}{\pgf@xa-0.5\pgf@xc}%
    \fi
    \fi
}

\pgfaddtoshape{beshape}{
    \anchorlet{c}{center}
    \anchorlet{n}{north}
    \anchorlet{e}{east}
    \anchorlet{s}{south}
    \anchorlet{w}{west}
    \anchorlet{se}{south east}
    \anchorlet{sw}{south west}
    \anchorlet{ne}{north east}
    \anchorlet{nw}{north west}
    \anchorlet{wsw}{west south west}
    \anchorlet{wnw}{west north west}
    \anchorlet{ene}{east north east}
    \anchorlet{ese}{east south east}
    \anchorlet{nnw}{north north west}
    \anchorlet{nne}{north north east}
    \anchorlet{ssw}{south south west}
    \anchorlet{sse}{south south east}
}

\tikzset{
    /tikz/bekeys/.cd,
    height/.initial=1,
    width/.initial=0.5,
    color/.initial=O,
    direction/.initial=e,
    linestyle/.initial={linestyle, rounded corners = 0},
    nports/.initial=3,
    inverted/.initial=0,    
    sbe/.initial=0,
    angle/.initial=60,
    /tikz/be/.code={
        \pgfqkeys{/tikz/bekeys}{#1}%
        \tikzset{/tikz/bekeys/drawer/.expanded=%
            \if\pgfkeysvalueof{/tikz/bekeys/direction}e
                {\pgfkeysvalueof{/tikz/bekeys/width}}%
                {\pgfkeysvalueof{/tikz/bekeys/height}}%
                {0}%
                {-90}%
            \fi
            \if\pgfkeysvalueof{/tikz/bekeys/direction}w
                {\pgfkeysvalueof{/tikz/bekeys/width}}%
                {\pgfkeysvalueof{/tikz/bekeys/height}}%
                {0}%
                {90}%
            \fi
            \if\pgfkeysvalueof{/tikz/bekeys/direction}n
                {\pgfkeysvalueof{/tikz/bekeys/width}}%
                {\pgfkeysvalueof{/tikz/bekeys/height}}%
                {1}%
                {0}%
            \fi
            \if\pgfkeysvalueof{/tikz/bekeys/direction}s
                {\pgfkeysvalueof{/tikz/bekeys/width}}%
                {\pgfkeysvalueof{/tikz/bekeys/height}}%
                {1}%
                {180}%
            \fi
            {\pgfkeysvalueof{/tikz/bekeys/color}}%
            {\pgfkeysvalueof{/tikz/bekeys/linestyle}}%
            {\pgfkeysvalueof{/tikz/bekeys/sbe}}%
            {\pgfkeysvalueof{/tikz/bekeys/angle}}%
        }
    },
    /tikz/bekeys/drawer/.code n args={8}{%
        \tikzset{
            beshape,
            #6,
            #5,
            minimum height=
            \ifnum#3>0    
            #1*\NODESIZE
            \else
            #2*\NODESIZE
            \fi
            ,minimum width=
            \ifnum#3>0
            #2*\NODESIZE
            \else
            #1*\NODESIZE
            \fi
            ,append after command={
                \pgfextra{
                    \let\bdr=\tikzlastnode%
                    \node[trapezium, line width = \NODETHICKNESS, minimum height=#1*\NODESIZE, minimum width=#2*\NODESIZE, trapezium stretches=true, rotate=#4, trapezium angle=70, inner sep=0.001mm, #5, #6] at (\bdr) (trap) {};

                    \node[rectangle, line width = \NODETHICKNESS, minimum height=#1*\NODESIZE, minimum width=#2*\NODESIZE, anchor=north, rotate=#4, #5, #6] at (trap.south) (r1) {};

                    \tikzmath{coordinate \C;
                    \C = (trap.top left corner)-(trap.top right corner);
                    \distAB = sqrt((\Cx)^2+(\Cy)^2);
                    }

                    \node[rectangle, line width = \NODETHICKNESS, minimum height=#1*\NODESIZE, minimum width=\distAB, anchor=south, rotate=#4, #5, #6, red] at (trap.north) (r2) {};

					\draw[#5, #6] (trap.top left corner) to (r2.north west) to (r2.north east) to (trap.top right corner) to (trap.bottom right corner) to (r1.south east) to (r1.south west) to (r1.north west) to cycle;

                }
            }
        }
    },
}
\pgfdeclareshape{couplershape}{	

	\inheritsavedanchors[from=basic] 

	\inheritanchor[from=basic]{center}
	\inheritanchor[from=basic]{mid}		
	\inheritanchor[from=basic]{base}	
	\inheritanchor[from=basic]{north}
	\inheritanchor[from=basic]{south}		
	\inheritanchor[from=basic]{west}		
	\inheritanchor[from=basic]{mid west}				
	\inheritanchor[from=basic]{base west}		
	\inheritanchor[from=basic]{north west}		
	\inheritanchor[from=basic]{south west}		
	\inheritanchor[from=basic]{east}
	\inheritanchor[from=basic]{mid east}
	\inheritanchor[from=basic]{base east}	
	\inheritanchor[from=basic]{north east}
	\inheritanchor[from=basic]{south east}
	\inheritanchor[from=basic]{west south west}%
	\inheritanchor[from=basic]{west north west}%
	\inheritanchor[from=basic]{east north east}%
	\inheritanchor[from=basic]{east south east}%
	\inheritanchor[from=basic]{north north west}%
	\inheritanchor[from=basic]{north north east}%
	\inheritanchor[from=basic]{south south west}%
	\inheritanchor[from=basic]{south south east}%
	
	\inheritanchorborder[from=basic]	
	\inheritbackgroundpath[from=basic]
}

\pgfaddtoshape{couplershape}{
	\anchorlet{in}{center}		
	\anchorlet{in0}{center}
	\anchorlet{out}{center}
	\anchorlet{out0}{center}
	\anchorlet{c}{center}		
	\anchorlet{n}{north}
	\anchorlet{e}{east}
	\anchorlet{s}{south}
	\anchorlet{w}{west}			
	\anchorlet{se}{south east}
	\anchorlet{sw}{south west}
	\anchorlet{ne}{north east}
	\anchorlet{nw}{north west}
	\anchorlet{wsw}{west south west}
	\anchorlet{wnw}{west north west}
	\anchorlet{ene}{east north east}
	\anchorlet{ese}{east south east}
	\anchorlet{nnw}{north north west}
	\anchorlet{nne}{north north east}
	\anchorlet{ssw}{south south west}
	\anchorlet{sse}{south south east}
}

\tikzset{
	/tikz/couplerkeys/.cd,
	size/.initial=0.2,
	color/.initial=O,
	rotation/.initial=0,
	heightwidthratio/.initial=0.5,
	/tikz/coupler/.code={
		\pgfqkeys{/tikz/couplerkeys}{#1}%
		\tikzset{/tikz/couplerkeys/drawer/.expanded=%
			{\pgfkeysvalueof{/tikz/couplerkeys/size}}%
			{\pgfkeysvalueof{/tikz/couplerkeys/color}}%
			{\pgfkeysvalueof{/tikz/couplerkeys/rotation}}%
			{\pgfkeysvalueof{/tikz/couplerkeys/heightwidthratio}}%
		}
	},
	/tikz/couplerkeys/drawer/.code n args={4}{%
		\tikzset{
			couplershape,
			minimum height=#1*\NODESIZE
			\ifnum#3<1
				\ifnum#3>-1
					*#4
				\fi
			\fi
			,minimum width=#1*\NODESIZE
			\ifnum#3<91
				\ifnum#3>89
					*#4
				\fi
			\fi
			\ifnum#3<-89
				\ifnum#3>-91
					*#4
				\fi
			\fi
			,#2,
			append after command={
				\pgfextra{\let\bdr=\tikzlastnode%
				\node[ellipse, fill, #2, rotate=#3, outer sep = 0, minimum width=#1*\NODESIZE, minimum height=#1*#4*\NODESIZE] at (\bdr.center){};
				}
			}
		}
	},
}
\pgfdeclareshape{fibershape}{
	\savedmacro\direction{
		\edef\direction{\pgfkeysvalueof{/tikz/fiberkeys/direction}}%
	}
	\savedmacro\flip{
		\edef\flip{\pgfkeysvalueof{/tikz/fiberkeys/flip}}%
	}
	\saveddimen\minwidth{
		\pgfmathsetlength\pgf@x{\pgfshapeminwidth}%
	}
	\saveddimen\minheight{
		\pgfmathsetlength\pgf@x{\pgfshapeminheight}%
	}
	\inheritsavedanchors[from=basic] 
	
	\inheritanchor[from=basic]{center}
	\inheritanchor[from=basic]{mid}		
	\inheritanchor[from=basic]{base}	
	\inheritanchor[from=basic]{north}
	\inheritanchor[from=basic]{south}		
	\inheritanchor[from=basic]{west}		
	\inheritanchor[from=basic]{mid west}				
	\inheritanchor[from=basic]{base west}		
	\inheritanchor[from=basic]{north west}		
	\inheritanchor[from=basic]{south west}		
	\inheritanchor[from=basic]{east}
	\inheritanchor[from=basic]{mid east}
	\inheritanchor[from=basic]{base east}	
	\inheritanchor[from=basic]{north east}
	\inheritanchor[from=basic]{south east}
	\inheritanchor[from=basic]{west south west}%
	\inheritanchor[from=basic]{west north west}%
	\inheritanchor[from=basic]{east north east}%
	\inheritanchor[from=basic]{east south east}%
	\inheritanchor[from=basic]{north north west}%
	\inheritanchor[from=basic]{north north east}%
	\inheritanchor[from=basic]{south south west}%
	\inheritanchor[from=basic]{south south east}%
	
	\inheritanchorborder[from=basic]	
	\inheritbackgroundpath[from=basic]

    \pgfutil@g@addto@macro\pgf@sh@s@fibershape{%
        \pgfutil@ifundefined{pgf@anchor@fibershape@in0}{
	        \expandafter\xdef\csname pgf@anchor@fibershape@in0\endcsname{%
	            \noexpand\fibershape@port{0}
	        }%
	    }{}%
        \pgfutil@ifundefined{pgf@anchor@fibershape@in}{
	        \expandafter\xdef\csname pgf@anchor@fibershape@in\endcsname{%
	            \noexpand\fibershape@port{0}
	        }%
	    }{}%
        \pgfutil@ifundefined{pgf@anchor@fibershape@out0}{
	        \expandafter\xdef\csname pgf@anchor@fibershape@out0\endcsname{%
	            \noexpand\fibershape@port{1}
	        }%
	    }{}%
        \pgfutil@ifundefined{pgf@anchor@fibershape@out}{
	        \expandafter\xdef\csname pgf@anchor@fibershape@out\endcsname{%
	            \noexpand\fibershape@port{1}
	        }%
	    }{}%
	}
}

\def\fibershape@port#1{
    \northeast	

    \ifnum#1=0	
	    \if\direction\direce
			\pgf@x=-\pgf@x
	    	\if\flip\flipfalse
		    	\pgf@y=-\pgf@y
		    \else
		    	\pgf@y=\pgf@y
		    \fi
		\fi
	    \if\direction\direcw
			\pgf@x=\pgf@x
	    	\if\flip\flipfalse
		    	\pgf@y=-\pgf@y
		    \else
		    	\pgf@y=\pgf@y
		    \fi
		\fi
	    \if\direction\direcn
			\pgf@y=-\pgf@y
	    	\if\flip\flipfalse
		    	\pgf@x=-\pgf@x
		    \else
		    	\pgf@x=\pgf@x
		    \fi
		\fi
	    \if\direction\direcs
			\pgf@y=\pgf@y
	    	\if\flip\flipfalse
		    	\pgf@x=-\pgf@x
		    \else
		    	\pgf@x=\pgf@x
		    \fi
		\fi
	\else	
	    \if\direction\direce
			\pgf@x=\pgf@x
	    	\if\flip\flipfalse
		    	\pgf@y=-\pgf@y
		    \else
		    	\pgf@y=\pgf@y
		    \fi
		\fi
	    \if\direction\direcw
			\pgf@x=-\pgf@x
	    	\if\flip\flipfalse
		    	\pgf@y=-\pgf@y
		    \else
		    	\pgf@y=\pgf@y
		    \fi
		\fi
	    \if\direction\direcn
			\pgf@y=\pgf@y
	    	\if\flip\flipfalse
		    	\pgf@x=-\pgf@x
		    \else
		    	\pgf@x=\pgf@x
		    \fi
		\fi
	    \if\direction\direcs
			\pgf@y=-\pgf@y
	    	\if\flip\flipfalse
		    	\pgf@x=-\pgf@x
		    \else
		    	\pgf@x=\pgf@x
		    \fi
		\fi
	\fi
}

\pgfaddtoshape{fibershape}{
	\anchorlet{c}{center}		
	\anchorlet{n}{north}
	\anchorlet{e}{east}
	\anchorlet{s}{south}
	\anchorlet{w}{west}			
	\anchorlet{se}{south east}
	\anchorlet{sw}{south west}
	\anchorlet{ne}{north east}
	\anchorlet{nw}{north west}
	\anchorlet{wsw}{west south west}
	\anchorlet{wnw}{west north west}
	\anchorlet{ene}{east north east}
	\anchorlet{ese}{east south east}
	\anchorlet{nnw}{north north west}
	\anchorlet{nne}{north north east}
	\anchorlet{ssw}{south south west}
	\anchorlet{sse}{south south east}
}

\tikzset{
	/tikz/fiberkeys/.cd,
	size/.initial=1,
	color/.initial=C0,
	direction/.initial=e,
	linestyle/.initial={linestyle},
	flip/.initial={0},
	drawbase/.initial={1},
	/tikz/fiber/.code={
		\pgfqkeys{/tikz/fiberkeys}{#1}%
		\tikzset{/tikz/fiberkeys/drawer/.expanded=%
			{\pgfkeysvalueof{/tikz/fiberkeys/direction}}%
			{\pgfkeysvalueof{/tikz/fiberkeys/size}}%
			{\pgfkeysvalueof{/tikz/fiberkeys/color}}%
			{\pgfkeysvalueof{/tikz/fiberkeys/linestyle}}%
			\if\pgfkeysvalueof{/tikz/fiberkeys/direction}e
				{a}%
				\if\pgfkeysvalueof{/tikz/fiberkeys/flip}0
					{south}%
				\else
					{north}%
				\fi
				{\pgfkeysvalueof{/tikz/fiberkeys/size}}
				{\pgfkeysvalueof{/tikz/fiberkeys/size} * 0.5}
			\fi
			\if\pgfkeysvalueof{/tikz/fiberkeys/direction}w
				{a}%
				\if\pgfkeysvalueof{/tikz/fiberkeys/flip}0
					{south}%
				\else
					{north}%
				\fi
				{\pgfkeysvalueof{/tikz/fiberkeys/size}}
				{\pgfkeysvalueof{/tikz/fiberkeys/size} * 0.5}
			\fi
			\if\pgfkeysvalueof{/tikz/fiberkeys/direction}n
				{b}%
				\if\pgfkeysvalueof{/tikz/fiberkeys/flip}0
					{west}%
				\else
					{east}%
				\fi
				{\pgfkeysvalueof{/tikz/fiberkeys/size} * 0.5}
				{\pgfkeysvalueof{/tikz/fiberkeys/size}}
			\fi
			\if\pgfkeysvalueof{/tikz/fiberkeys/direction}s
				{b}%
				\if\pgfkeysvalueof{/tikz/fiberkeys/flip}0
					{west}%
				\else
					{east}%
				\fi
				{\pgfkeysvalueof{/tikz/fiberkeys/size} * 0.5}
				{\pgfkeysvalueof{/tikz/fiberkeys/size}}
			\fi
			{\pgfkeysvalueof{/tikz/fiberkeys/drawbase}}%
		}
	},
	/tikz/fiberkeys/drawer/.code n args={9}{%
		\tikzset{
			fibershape,
			minimum width=#7*\NODESIZE,
			minimum height=#8*\NODESIZE,
			append after command={
				\pgfextra{\let\bdr=\tikzlastnode%
				\if#5a	
					\ifnum#9>0
						\draw[#3, #4] (\bdr.#6 west) to (\bdr.#6 east) {};
					\fi
					\node[draw=#3, #4, circle, minimum size=#2*0.5*\NODESIZE, anchor=#6] at ([xshift=-0.1*#2*\NODESIZE]\bdr.#6) () {};
					\node[draw=#3, #4, circle, minimum size=#2*0.5*\NODESIZE, anchor=#6] at (\bdr.#6) () {};
					\node[draw=#3, #4, circle, minimum size=#2*0.5*\NODESIZE, anchor=#6] at ([xshift=0.1*#2*\NODESIZE]\bdr.#6) () {};
				\fi
				\if#5b	
					\ifnum#9>0
						\draw[#3, #4] (\bdr.north #6) to (\bdr.south #6) {};
					\fi
					\node[draw=#3, #4, circle, minimum size=#2*0.5*\NODESIZE, anchor=#6] at ([yshift=0.1*#2*\NODESIZE]\bdr.#6) () {};
					\node[draw=#3, #4, circle, minimum size=#2*0.5*\NODESIZE, anchor=#6] at (\bdr.#6) () {};
					\node[draw=#3, #4, circle, minimum size=#2*0.5*\NODESIZE, anchor=#6] at ([yshift=-0.1*#2*\NODESIZE]\bdr.#6) () {};
				\fi
				}
			}
		}
	},
}
\newcount\portcount

\pgfdeclareshape{fiberswitchshape}{
	\savedmacro\nin{
		\edef\nin{\pgfkeysvalueof{/tikz/fiberswitchkeys/nin}}%
	}
	\savedmacro\nout{
		\edef\nout{\pgfkeysvalueof{/tikz/fiberswitchkeys/nout}}%
	}
	\savedmacro\direction{
		\edef\direction{\pgfkeysvalueof{/tikz/fiberswitchkeys/direction}}%
	}
	\inheritsavedanchors[from=basic] 
	
	\inheritanchor[from=basic]{center}
	\inheritanchor[from=basic]{mid}		
	\inheritanchor[from=basic]{base}	
	\inheritanchor[from=basic]{north}
	\inheritanchor[from=basic]{south}		
	\inheritanchor[from=basic]{west}		
	\inheritanchor[from=basic]{mid west}				
	\inheritanchor[from=basic]{base west}		
	\inheritanchor[from=basic]{north west}		
	\inheritanchor[from=basic]{south west}		
	\inheritanchor[from=basic]{east}
	\inheritanchor[from=basic]{mid east}
	\inheritanchor[from=basic]{base east}	
	\inheritanchor[from=basic]{north east}
	\inheritanchor[from=basic]{south east}
	\inheritanchor[from=basic]{west south west}%
	\inheritanchor[from=basic]{west north west}%
	\inheritanchor[from=basic]{east north east}%
	\inheritanchor[from=basic]{east south east}%
	\inheritanchor[from=basic]{north north west}%
	\inheritanchor[from=basic]{north north east}%
	\inheritanchor[from=basic]{south south west}%
	\inheritanchor[from=basic]{south south east}%
	
	\inheritanchorborder[from=basic]	
	\inheritbackgroundpath[from=basic]

    \pgfutil@g@addto@macro\pgf@sh@s@fiberswitchshape{%
        \pgfmathsetcount{\portcount}{0}
        \pgfmathloop%
        \ifnum\the\portcount<\nin
	        \pgfutil@ifundefined{pgf@anchor@fiberswitchshape@in\the\portcount}{
		        \expandafter\xdef\csname pgf@anchor@fiberswitchshape@in\the\portcount\endcsname{%
		            \noexpand\fiberswitchshape@port[\the\portcount]{0}
		        }%
		    }{}%
	        \ifnum\the\portcount=0
    		    \pgfutil@ifundefined{pgf@anchor@fiberswitchshape@in}{%
		        \expandafter\xdef\csname pgf@anchor@fiberswitchshape@in\endcsname{%
		            \noexpand\fiberswitchshape@port[\the\portcount]{0}
		        }%
		        }{}%
		    \fi
	        \pgfmathaddtocount{\portcount}{1}	
	        \repeatpgfmathloop
        \pgfmathsetcount{\portcount}{0}
        \pgfmathloop%
    	\ifnum\the\portcount<\nout
	        \pgfutil@ifundefined{pgf@anchor@fiberswitchshape@out\the\portcount}{%
		        \expandafter\xdef\csname pgf@anchor@fiberswitchshape@out\the\portcount\endcsname{%
		            \noexpand\fiberswitchshape@port[\the\portcount]{1}
		        }%
		    }{}%
	        \ifnum\the\portcount=0
    		    \pgfutil@ifundefined{pgf@anchor@fiberswitchshape@out}{%
		        \expandafter\xdef\csname pgf@anchor@fiberswitchshape@out\endcsname{%
		            \noexpand\fiberswitchshape@port[\the\portcount]{1}
		        }%
		        }{}%
		    \fi
	        \pgfmathaddtocount{\portcount}{1}	
	        \repeatpgfmathloop
	}
}

\def\fiberswitchshape@port[#1]#2{
    \northeast \pgf@xa=\pgf@x \pgf@ya=\pgf@y
    \southwest \pgf@xb=\pgf@x \pgf@yb=\pgf@y
    
    \ifnum#2=0	
	    \if\direction\direce	
	    	\pgf@x=\pgf@xb
		    \pgf@yc=\pgf@ya \advance\pgf@yc by -\pgf@yb	
		    \pgfmathsetlength{\pgf@y}{\pgf@ya-(#1 + 0.5)*(\pgf@yc/\nin)}%
	    \fi
	    \if\direction\direcw
	    	\pgf@x=\pgf@xa
		    \pgf@yc=\pgf@ya \advance\pgf@yc by -\pgf@yb	
		    \pgfmathsetlength{\pgf@y}{\pgf@ya-(#1 + 0.5)*(\pgf@yc/\nin)}%
	    \fi
	    \if\direction\direcn
	    	\pgf@y=\pgf@yb
		    \pgf@xc=\pgf@xa \advance\pgf@xc by -\pgf@xb	
		    \pgfmathsetlength{\pgf@x}{\pgf@xb+(#1 + 0.5)*(\pgf@xc/\nin)}%
	    \fi
	    \if\direction\direcs
	    	\pgf@y=\pgf@ya
		    \pgf@xc=\pgf@xa \advance\pgf@xc by -\pgf@xb	
		    \pgfmathsetlength{\pgf@x}{\pgf@xb+(#1 + 0.5)*(\pgf@xc/\nin)}%
	    \fi
	\else	
	    \if\direction\direce	
	    	\pgf@x=\pgf@xa
		    \pgf@yc=\pgf@ya \advance\pgf@yc by -\pgf@yb	
		    \pgfmathsetlength{\pgf@y}{\pgf@ya-(#1 + 0.5)*(\pgf@yc/\nout)}%
	    \fi
	    \if\direction\direcw
	    	\pgf@x=\pgf@xb
		    \pgf@yc=\pgf@ya \advance\pgf@yc by -\pgf@yb	
		    \pgfmathsetlength{\pgf@y}{\pgf@ya-(#1 + 0.5)*(\pgf@yc/\nout)}%
	    \fi
	    \if\direction\direcn
	    	\pgf@y=\pgf@ya
		    \pgf@xc=\pgf@xa \advance\pgf@xc by -\pgf@xb	
		    \pgfmathsetlength{\pgf@x}{\pgf@xb+(#1 + 0.5)*(\pgf@xc/\nout)}%
	    \fi
	    \if\direction\direcs
	    	\pgf@y=\pgf@yb
		    \pgf@xc=\pgf@xa \advance\pgf@xc by -\pgf@xb	
		    \pgfmathsetlength{\pgf@x}{\pgf@xb+(#1 + 0.5)*(\pgf@xc/\nout)}%
	    \fi
	\fi
}

\pgfaddtoshape{fiberswitchshape}{
	\anchorlet{c}{center}		
	\anchorlet{n}{north}
	\anchorlet{e}{east}
	\anchorlet{s}{south}
	\anchorlet{w}{west}			
	\anchorlet{se}{south east}
	\anchorlet{sw}{south west}
	\anchorlet{ne}{north east}
	\anchorlet{nw}{north west}
	\anchorlet{wsw}{west south west}
	\anchorlet{wnw}{west north west}
	\anchorlet{ene}{east north east}
	\anchorlet{ese}{east south east}
	\anchorlet{nnw}{north north west}
	\anchorlet{nne}{north north east}
	\anchorlet{ssw}{south south west}
	\anchorlet{sse}{south south east}
}

\tikzset{
	/tikz/fiberswitchkeys/.cd,
	size/.initial=1,
	color/.initial=O,
	direction/.initial=e,
	linestyle/.initial={linestyle, inner sep=0.5mm},
	nin/.initial=1,	
	nout/.initial=3,
	/tikz/fiberswitch/.code={
		\pgfqkeys{/tikz/fiberswitchkeys}{#1}%
		\tikzset{/tikz/fiberswitchkeys/drawer/.expanded=%
			{\pgfkeysvalueof{/tikz/fiberswitchkeys/size}}%
			{\pgfkeysvalueof{/tikz/fiberswitchkeys/color}}%
			{\pgfkeysvalueof{/tikz/fiberswitchkeys/linestyle}}%
			{\pgfkeysvalueof{/tikz/fiberswitchkeys/nout}}%
			\if\pgfkeysvalueof{/tikz/fiberswitchkeys/direction}e
				{0}%
			\fi
			\if\pgfkeysvalueof{/tikz/fiberswitchkeys/direction}w
				{0}%
			\fi
			\if\pgfkeysvalueof{/tikz/fiberswitchkeys/direction}n
				{1}%
			\fi
			\if\pgfkeysvalueof{/tikz/fiberswitchkeys/direction}s
				{1}%
			\fi
			{\pgfkeysvalueof{/tikz/fiberswitchkeys/direction}}%
		}
	},
	/tikz/fiberswitchkeys/drawer/.code n args={6}{%
		\tikzset{
			fiberswitchshape,
			draw,
			minimum height = #1*\NODESIZE,
			minimum width = #1*\NODESIZE,
			#2,
			#3,
			append after command={
				\pgfextra{\let\bdr=\tikzlastnode%
						\ifnum#5>0
							\node[coordinate] at ($(\bdr.in)!0.25!(\bdr.out0 -| \bdr.in)$) (circlein){};
							\foreach \n [evaluate=\n as \nport using int(\n-1)] in {1,...,#4}{
								\node[coordinate] at ($(\bdr.out\nport)!0.25!(\bdr.in -| \bdr.out\nport)$) (circleout\nport){};
							}
						\else
							\node[coordinate] at ($(\bdr.in)!0.25!(\bdr.out0 |- \bdr.in)$) (circlein){};
							\foreach \n [evaluate=\n as \nport using int(\n-1)] in {1,...,#4}{
								\node[coordinate] at ($(\bdr.out\nport)!0.25!(\bdr.in |- \bdr.out\nport)$) (circleout\nport){};
							}
						\fi

						\draw[---, #2, #3, fill] (\bdr.in) to (circlein) circle (0.05);
						\foreach \n [evaluate=\n as \nport using int(\n-1)] in {1,...,#4}{
							\draw[---, #2, #3, fill] (\bdr.out\nport) to (circleout\nport) circle (0.05);
						}

						\tikzmath{
							int \nportmax;
							\nportmax = int(#4-1);
						}
						\draw[---, #2, #3] (circlein) to (circleout0){};
						\if#6e
							\draw[-->, #2, #3, looseness=0.8] ($(circlein)!0.6!(circleout0)$) to [out=-60, in=60]($(circlein)!0.6!(circleout\nportmax)$) {};
						\fi
						\if#6w
							\draw[-->, #2, #3, looseness=0.8] ($(circlein)!0.6!(circleout0)$) to [out=-120, in=120]($(circlein)!0.6!(circleout\nportmax)$) {};
						\fi
						\if#6s
							\draw[-->, #2, #3, looseness=0.8] ($(circlein)!0.6!(circleout0)$) to [out=-30, in=-150]($(circlein)!0.6!(circleout\nportmax)$) {};
						\fi
						\if#6n
							\draw[-->, #2, #3, looseness=0.8] ($(circlein)!0.6!(circleout0)$) to [out=30, in=150]($(circlein)!0.6!(circleout\nportmax)$) {};
						\fi

				}
			}
		}
	},
}
\pgfdeclareshape{filtershape}{
	\savedmacro\direction{
		\edef\direction{\pgfkeysvalueof{/tikz/filterkeys/direction}}%
	}
	\saveddimen\minwidth{
		\pgfmathsetlength\pgf@x{\pgfshapeminwidth}%
	}
	\saveddimen\minheight{
		\pgfmathsetlength\pgf@x{\pgfshapeminheight}%
	}
	\inheritsavedanchors[from=basic]

	\inheritanchor[from=basic]{center}
	\inheritanchor[from=basic]{mid}
	\inheritanchor[from=basic]{base}
	\inheritanchor[from=basic]{north}
	\inheritanchor[from=basic]{south}
	\inheritanchor[from=basic]{west}
	\inheritanchor[from=basic]{mid west}
	\inheritanchor[from=basic]{base west}
	\inheritanchor[from=basic]{north west}
	\inheritanchor[from=basic]{south west}
	\inheritanchor[from=basic]{east}
	\inheritanchor[from=basic]{mid east}
	\inheritanchor[from=basic]{base east}
	\inheritanchor[from=basic]{north east}
	\inheritanchor[from=basic]{south east}
	\inheritanchor[from=basic]{west south west}%
	\inheritanchor[from=basic]{west north west}%
	\inheritanchor[from=basic]{east north east}%
	\inheritanchor[from=basic]{east south east}%
	\inheritanchor[from=basic]{north north west}%
	\inheritanchor[from=basic]{north north east}%
	\inheritanchor[from=basic]{south south west}%
	\inheritanchor[from=basic]{south south east}%

	\inheritanchorborder[from=basic]
	\inheritbackgroundpath[from=basic]

	\pgfutil@g@addto@macro\pgf@sh@s@filtershape{%
		\pgfutil@ifundefined{pgf@anchor@filtershape@in0}{
			\expandafter\xdef\csname pgf@anchor@filtershape@in0\endcsname{%
				\noexpand\filtershape@port{0}
			}%
		}{}%
		\pgfutil@ifundefined{pgf@anchor@filtershape@in}{
			\expandafter\xdef\csname pgf@anchor@filtershape@in\endcsname{%
				\noexpand\filtershape@port{0}
			}%
		}{}%
		\pgfutil@ifundefined{pgf@anchor@filtershape@out0}{
			\expandafter\xdef\csname pgf@anchor@filtershape@out0\endcsname{%
				\noexpand\filtershape@port{1}
			}%
		}{}%
		\pgfutil@ifundefined{pgf@anchor@filtershape@out}{
			\expandafter\xdef\csname pgf@anchor@filtershape@out\endcsname{%
				\noexpand\filtershape@port{1}
			}%
		}{}%
	}
}

\def\filtershape@port#1{
	\northeast	

	\ifnum#1=0	
		\if\direction\direce
			\pgf@x=-\pgf@x
			\pgf@ya= \pgf@y
			\pgfmathsetlength{\pgf@y}{\pgf@ya-0.5*\minheight}%
		\fi
		\if\direction\direcw
			\pgf@x=\pgf@x
			\pgf@ya= \pgf@y
			\pgfmathsetlength{\pgf@y}{\pgf@ya-0.5*\minheight}%
		\fi
		\if\direction\direcn
			\pgf@y=-\pgf@y
			\pgf@xa=\pgf@x
			\pgfmathsetlength{\pgf@x}{\pgf@xa-0.5*\minwidth}%
		\fi
		\if\direction\direcs
			\pgf@y=\pgf@y
			\pgf@xa= \pgf@x
			\pgfmathsetlength{\pgf@x}{\pgf@xa-0.5*\minwidth}%
		\fi
	\else	
		\if\direction\direce
			\pgf@x=\pgf@x
			\pgf@ya= \pgf@y
			\pgfmathsetlength{\pgf@y}{\pgf@ya-0.5*\minheight}%
		\fi
		\if\direction\direcw
			\pgf@x=-\pgf@x
			\pgf@ya= \pgf@y
			\pgfmathsetlength{\pgf@y}{\pgf@ya-0.5*\minheight}%
		\fi
		\if\direction\direcn
			\pgf@y=\pgf@y
			\pgf@xa= \pgf@x
			\pgfmathsetlength{\pgf@x}{\pgf@xa-0.5*\minwidth}%
		\fi
		\if\direction\direcs
			\pgf@y=-\pgf@y
			\pgf@xa= \pgf@x
			\pgfmathsetlength{\pgf@x}{\pgf@xa-0.5*\minwidth}%
		\fi
	\fi
}

\pgfaddtoshape{filtershape}{
	\anchorlet{c}{center}
	\anchorlet{n}{north}
	\anchorlet{e}{east}
	\anchorlet{s}{south}
	\anchorlet{w}{west}
	\anchorlet{se}{south east}
	\anchorlet{sw}{south west}
	\anchorlet{ne}{north east}
	\anchorlet{nw}{north west}
	\anchorlet{wsw}{west south west}
	\anchorlet{wnw}{west north west}
	\anchorlet{ene}{east north east}
	\anchorlet{ese}{east south east}
	\anchorlet{nnw}{north north west}
	\anchorlet{nne}{north north east}
	\anchorlet{ssw}{south south west}
	\anchorlet{sse}{south south east}
}

\pgfmathsetmacro{\WSSSINEHEIGHT}{0.06}

\tikzset{
	/tikz/filterkeys/.cd,
	size/.initial=0.5,
	color/.initial=O,
	direction/.initial=e,
	linestyle/.initial={linestyle},
	fillgradient/.initial=O,
	/tikz/filter/.code={
			\pgfqkeys{/tikz/filterkeys}{#1}%
			\tikzset{/tikz/filterkeys/drawer/.expanded=%
					{\pgfkeysvalueof{/tikz/filterkeys/direction}}%
					{\pgfkeysvalueof{/tikz/filterkeys/size}}%
					{\pgfkeysvalueof{/tikz/filterkeys/color}}%
					{\pgfkeysvalueof{/tikz/filterkeys/linestyle}}%
					{\pgfkeysvalueof{/tikz/filterkeys/fillgradient}}%
			}
		},
	/tikz/filterkeys/drawer/.code n args={5}{%
			\tikzset{
				filtershape,
				minimum height=#2*\NODESIZE,
				minimum width=#2*\NODESIZE,
				#3,
				#4,
				draw,
				append after command={
						\pgfextra{\let\bdr=\tikzlastnode%
							\node[#5, fit=(\bdr.nw)(\bdr.se)] (boxgradient){};

							\node[coordinate] at (\bdr.wnw -| \bdr.nnw) (hl){};
							\node[coordinate] at (\bdr.ene -| \bdr.nne) (hr){};
							\draw[#3,---, rounded corners = 0] (hl) sin ($(hl)!0.25!(hr) + (0,0.002*#2*\NODESIZE)$) cos ($(hl)!0.5!(hr)$) sin ($(hl)!0.75!(hr) + (0,-0.002*#2*\NODESIZE)$) cos (hr);

							\node[coordinate] at (\bdr.w -| \bdr.nnw) (ml){};
							\node[coordinate] at (\bdr.e -| \bdr.nne) (mr){};
							\draw[#3,---, rounded corners = 0] (ml) sin ($(ml)!0.25!(mr) + (0,0.002*#2*\NODESIZE)$) cos ($(ml)!0.5!(mr)$) sin ($(ml)!0.75!(mr) + (0,-0.002*#2*\NODESIZE)$) cos (mr);

							\node[coordinate] at (\bdr.wsw -| \bdr.nnw) (ll){};
							\node[coordinate] at (\bdr.ese -| \bdr.nne) (lr){};
							\draw[#3,---, rounded corners = 0] (ll) sin ($(ll)!0.25!(lr) + (0,0.002*#2*\NODESIZE)$) cos ($(ll)!0.5!(lr)$) sin ($(ll)!0.75!(lr) + (0,-0.002*#2*\NODESIZE)$) cos (lr);

							\draw[#3, ---] ($(hl)!0.25!(hr) - (0,0.002*#2*\NODESIZE)$) -- ($(hl)!0.75!(hr) + (0,0.002*#2*\NODESIZE)$);
							\draw[#3, ---] ($(ll)!0.25!(lr) - (0,0.002*#2*\NODESIZE)$) -- ($(ll)!0.75!(lr) + (0,0.002*#2*\NODESIZE)$);
						}
					}
			}
		},
}
\newcount\portcount

\pgfdeclareshape{lensshape}{
	\savedmacro\nport{
		\edef\nport{\pgfkeysvalueof{/tikz/lenskeys/nport}}%
	}
	\inheritsavedanchors[from=basic] 
	
	\inheritanchor[from=basic]{center}
	\inheritanchor[from=basic]{mid}		
	\inheritanchor[from=basic]{base}	
	\inheritanchor[from=basic]{north}
	\inheritanchor[from=basic]{south}		
	\inheritanchor[from=basic]{west}		
	\inheritanchor[from=basic]{mid west}				
	\inheritanchor[from=basic]{base west}		
	\inheritanchor[from=basic]{north west}		
	\inheritanchor[from=basic]{south west}		
	\inheritanchor[from=basic]{east}
	\inheritanchor[from=basic]{mid east}
	\inheritanchor[from=basic]{base east}	
	\inheritanchor[from=basic]{north east}
	\inheritanchor[from=basic]{south east}
	\inheritanchor[from=basic]{west south west}%
	\inheritanchor[from=basic]{west north west}%
	\inheritanchor[from=basic]{east north east}%
	\inheritanchor[from=basic]{east south east}%
	\inheritanchor[from=basic]{north north west}%
	\inheritanchor[from=basic]{north north east}%
	\inheritanchor[from=basic]{south south west}%
	\inheritanchor[from=basic]{south south east}%
	
	\inheritanchorborder[from=basic]	
	\inheritbackgroundpath[from=basic]

    \pgfutil@g@addto@macro\pgf@sh@s@lensshape{%
        \pgfmathsetcount{\portcount}{0}
        \pgfmathloop%
        \ifnum\the\portcount<\nport
	        \pgfutil@ifundefined{pgf@anchor@lensshape@in\the\portcount}{
		        \expandafter\xdef\csname pgf@anchor@lensshape@in\the\portcount\endcsname{%
		            \noexpand\lensshape@port[\the\portcount]
		        }%
		    }{}%
	        \ifnum\the\portcount=0
    		    \pgfutil@ifundefined{pgf@anchor@lensshape@in}{%
		        \expandafter\xdef\csname pgf@anchor@lensshape@in\endcsname{%
		            \noexpand\lensshape@port[\the\portcount]
		        }%
		        }{}%
		    \fi
		    \pgfutil@ifundefined{pgf@anchor@lensshape@out\the\portcount}{%
		        \expandafter\xdef\csname pgf@anchor@lensshape@out\the\portcount\endcsname{%
		            \noexpand\lensshape@port[\the\portcount]
		        }%
		    }{}%
	        \ifnum\the\portcount=0
    		    \pgfutil@ifundefined{pgf@anchor@lensshape@out}{%
		        \expandafter\xdef\csname pgf@anchor@lensshape@out\endcsname{%
		            \noexpand\lensshape@port[\the\portcount]
		        }%
		        }{}%
		    \fi
	        \pgfmathaddtocount{\portcount}{1}	
	        \repeatpgfmathloop
	}
}

\def\lensshape@port[#1]{
    \northeast \pgf@xa=\pgf@x \pgf@ya=\pgf@y
    \southwest \pgf@xb=\pgf@x \pgf@yb=\pgf@y
    
	\pgfmathsetlength{\pgf@x}{0.5*\pgf@xa + 0.5*\pgf@xb}%
    \pgf@yc=\pgf@ya \advance\pgf@yc by -\pgf@yb	
    \pgfmathsetlength{\pgf@y}{\pgf@ya-(#1 + 0.5)*(\pgf@yc/\nport)}%
}

\pgfaddtoshape{lensshape}{
	\anchorlet{c}{center}		
	\anchorlet{n}{north}
	\anchorlet{e}{east}
	\anchorlet{s}{south}
	\anchorlet{w}{west}			
	\anchorlet{se}{south east}
	\anchorlet{sw}{south west}
	\anchorlet{ne}{north east}
	\anchorlet{nw}{north west}
	\anchorlet{wsw}{west south west}
	\anchorlet{wnw}{west north west}
	\anchorlet{ene}{east north east}
	\anchorlet{ese}{east south east}
	\anchorlet{nnw}{north north west}
	\anchorlet{nne}{north north east}
	\anchorlet{ssw}{south south west}
	\anchorlet{sse}{south south east}
}

\tikzset{
	/tikz/lenskeys/.cd,
	height/.initial=1,
	width/.initial=0.3,	
	color/.initial=O,
	rotation/.initial=0,
	linestyle/.initial={linestyle, inner sep=0.5mm},
	nport/.initial=1,
	/tikz/lens/.code={
		\pgfqkeys{/tikz/lenskeys}{#1}%
		\tikzset{/tikz/lenskeys/drawer/.expanded=%
			{\pgfkeysvalueof{/tikz/lenskeys/height}}%
			{\pgfkeysvalueof{/tikz/lenskeys/width}}%
			{\pgfkeysvalueof{/tikz/lenskeys/color}}%
			{\pgfkeysvalueof{/tikz/lenskeys/linestyle}}%
			{\pgfkeysvalueof{/tikz/lenskeys/rotation}}%
		}
	},
	/tikz/lenskeys/drawer/.code n args={5}{%
		\tikzset{
			lensshape,
			rotate=#5,
			minimum height=#1*\NODESIZE,
			minimum width=#2*\NODESIZE,
			append after command={
				\pgfextra{\let\bdr=\tikzlastnode%
					\draw[---, #3, #4, rounded corners = 0] (\bdr.n) to [in=120+#5, out=240+#5] (\bdr.s) to [in=-60+#5, out=60+#5] (\bdr.n) -- cycle;
				}
			}
		}
	},
}

\newcount\portcount

\pgfdeclareshape{mirrorshape}{
	\savedmacro\nport{
		\edef\nport{\pgfkeysvalueof{/tikz/mirrorkeys/nport}}%
	}
	\inheritsavedanchors[from=basic] 
	
	\inheritanchor[from=basic]{center}
	\inheritanchor[from=basic]{mid}		
	\inheritanchor[from=basic]{base}	
	\inheritanchor[from=basic]{north}
	\inheritanchor[from=basic]{south}		
	\inheritanchor[from=basic]{west}		
	\inheritanchor[from=basic]{mid west}				
	\inheritanchor[from=basic]{base west}		
	\inheritanchor[from=basic]{north west}		
	\inheritanchor[from=basic]{south west}		
	\inheritanchor[from=basic]{east}
	\inheritanchor[from=basic]{mid east}
	\inheritanchor[from=basic]{base east}	
	\inheritanchor[from=basic]{north east}
	\inheritanchor[from=basic]{south east}
	\inheritanchor[from=basic]{west south west}%
	\inheritanchor[from=basic]{west north west}%
	\inheritanchor[from=basic]{east north east}%
	\inheritanchor[from=basic]{east south east}%
	\inheritanchor[from=basic]{north north west}%
	\inheritanchor[from=basic]{north north east}%
	\inheritanchor[from=basic]{south south west}%
	\inheritanchor[from=basic]{south south east}%
	
	\inheritanchorborder[from=basic]	
	\inheritbackgroundpath[from=basic]

    \pgfutil@g@addto@macro\pgf@sh@s@mirrorshape{%
        \pgfmathsetcount{\portcount}{0}
        \pgfmathloop%
        \ifnum\the\portcount<\nport
	        \pgfutil@ifundefined{pgf@anchor@mirrorshape@in\the\portcount}{
		        \expandafter\xdef\csname pgf@anchor@mirrorshape@in\the\portcount\endcsname{%
		            \noexpand\mirrorshape@port[\the\portcount]
		        }%
		    }{}%
	        \ifnum\the\portcount=0
    		    \pgfutil@ifundefined{pgf@anchor@mirrorshape@in}{%
		        \expandafter\xdef\csname pgf@anchor@mirrorshape@in\endcsname{%
		            \noexpand\mirrorshape@port[\the\portcount]
		        }%
		        }{}%
		    \fi
		    \pgfutil@ifundefined{pgf@anchor@mirrorshape@out\the\portcount}{%
		        \expandafter\xdef\csname pgf@anchor@mirrorshape@out\the\portcount\endcsname{%
		            \noexpand\mirrorshape@port[\the\portcount]
		        }%
		    }{}%
	        \ifnum\the\portcount=0
    		    \pgfutil@ifundefined{pgf@anchor@mirrorshape@out}{%
		        \expandafter\xdef\csname pgf@anchor@mirrorshape@out\endcsname{%
		            \noexpand\mirrorshape@port[\the\portcount]
		        }%
		        }{}%
		    \fi
	        \pgfmathaddtocount{\portcount}{1}	
	        \repeatpgfmathloop
	}
}

\def\mirrorshape@port[#1]{
    \northeast \pgf@xa=\pgf@x \pgf@ya=\pgf@y
    \southwest \pgf@xb=\pgf@x \pgf@yb=\pgf@y
    
	\pgf@x=\pgf@xb
    \pgf@yc=\pgf@ya \advance\pgf@yc by -\pgf@yb	
    \pgfmathsetlength{\pgf@y}{\pgf@ya-(#1 + 0.5)*(\pgf@yc/\nport)}%
}

\pgfaddtoshape{mirrorshape}{
	\anchorlet{c}{center}		
	\anchorlet{n}{north}
	\anchorlet{e}{east}
	\anchorlet{s}{south}
	\anchorlet{w}{west}			
	\anchorlet{se}{south east}
	\anchorlet{sw}{south west}
	\anchorlet{ne}{north east}
	\anchorlet{nw}{north west}
	\anchorlet{wsw}{west south west}
	\anchorlet{wnw}{west north west}
	\anchorlet{ene}{east north east}
	\anchorlet{ese}{east south east}
	\anchorlet{nnw}{north north west}
	\anchorlet{nne}{north north east}
	\anchorlet{ssw}{south south west}
	\anchorlet{sse}{south south east}
}

\tikzset{
	/tikz/mirrorkeys/.cd,
	height/.initial=1,
	width/.initial=0.15,	
	color/.initial=O,
	rotation/.initial=0,
	linestyle/.initial={linestyle, inner sep=0.5mm},
	nport/.initial=1,
	nlines/.initial=5,
	/tikz/mirror/.code={
		\pgfqkeys{/tikz/mirrorkeys}{#1}%
		\tikzset{/tikz/mirrorkeys/drawer/.expanded=%
			{\pgfkeysvalueof{/tikz/mirrorkeys/height}}%
			{\pgfkeysvalueof{/tikz/mirrorkeys/width}}%
			{\pgfkeysvalueof{/tikz/mirrorkeys/color}}%
			{\pgfkeysvalueof{/tikz/mirrorkeys/linestyle}}%
			{\pgfkeysvalueof{/tikz/mirrorkeys/rotation}}%
			{\pgfkeysvalueof{/tikz/mirrorkeys/nlines}}%
		}
	},
	/tikz/mirrorkeys/drawer/.code n args={6}{%
		\tikzset{
			mirrorshape,
			rotate=#5,
			minimum height=#1*\NODESIZE,
			minimum width=#2*\NODESIZE,
			append after command={
				\pgfextra{\let\bdr=\tikzlastnode%
					\draw[---, #3, #4] (\bdr.nw) to (\bdr.sw){};
					\foreach \nline [evaluate=\nline as \linepos using (\nline-0.8)/(#6-0.6)] in {1,...,#6}{
						\draw[---, #3, #4] ($(\bdr.nw)!\linepos!(\bdr.sw)$) to +(#5-45:#2*1.41421356237*\NODESIZE pt){};
					}
				}
			}
		}
	},
}
\pgfdeclareshape{muxshape}{
	\savedmacro\nports{
		\edef\nports{\pgfkeysvalueof{/tikz/muxkeys/nports}}%
	}
	\savedmacro\direction{
		\edef\direction{\pgfkeysvalueof{/tikz/muxkeys/direction}}%
	}
	\savedmacro\inverted{
		\edef\inverted{\pgfkeysvalueof{/tikz/muxkeys/inverted}}%
	}
	\savedmacro\ninports{
		\ifnum\inverted=0
			\edef\ninports{\nports}
		\else
			\edef\ninports{1}%
		\fi
	}
	\savedmacro\noutports{
		\ifnum\inverted=0
			\edef\noutports{1}%
		\else
			\edef\noutports{\nports}%
		\fi
	}
	\inheritsavedanchors[from=basic]

	\inheritanchor[from=basic]{center}
	\inheritanchor[from=basic]{mid}
	\inheritanchor[from=basic]{base}
	\inheritanchor[from=basic]{north}
	\inheritanchor[from=basic]{south}
	\inheritanchor[from=basic]{west}
	\inheritanchor[from=basic]{mid west}
	\inheritanchor[from=basic]{base west}
	\inheritanchor[from=basic]{north west}
	\inheritanchor[from=basic]{south west}
	\inheritanchor[from=basic]{east}
	\inheritanchor[from=basic]{mid east}
	\inheritanchor[from=basic]{base east}
	\inheritanchor[from=basic]{north east}
	\inheritanchor[from=basic]{south east}
	\inheritanchor[from=basic]{west south west}%
	\inheritanchor[from=basic]{west north west}%
	\inheritanchor[from=basic]{east north east}%
	\inheritanchor[from=basic]{east south east}%
	\inheritanchor[from=basic]{north north west}%
	\inheritanchor[from=basic]{north north east}%
	\inheritanchor[from=basic]{south south west}%
	\inheritanchor[from=basic]{south south east}%

	\inheritanchorborder[from=basic]
	\inheritbackgroundpath[from=basic]

	\pgfutil@g@addto@macro\pgf@sh@s@muxshape{%
		\pgfmathsetcount{\portcount}{0}
		\pgfmathloop%
		\ifnum\the\portcount<\nports
		\ifnum\the\portcount<\ninports
			\pgfutil@ifundefined{pgf@anchor@muxshape@in\the\portcount}{
				\expandafter\xdef\csname pgf@anchor@muxshape@in\the\portcount\endcsname{%
					\noexpand\muxshape@port[\the\portcount]{0}
				}%
			}{}%
			\ifnum\the\portcount=0
				\pgfutil@ifundefined{pgf@anchor@muxshape@in}{%
					\expandafter\xdef\csname pgf@anchor@muxshape@in\endcsname{%
						\noexpand\muxshape@port[\the\portcount]{0}
					}%
				}{}%
			\fi
		\fi
		\ifnum\the\portcount<\noutports
			\pgfutil@ifundefined{pgf@anchor@muxshape@out\the\portcount}{%
				\expandafter\xdef\csname pgf@anchor@muxshape@out\the\portcount\endcsname{%
					\noexpand\muxshape@port[\the\portcount]{1}
				}%
			}{}%
			\ifnum\the\portcount=0
				\pgfutil@ifundefined{pgf@anchor@muxshape@out}{%
					\expandafter\xdef\csname pgf@anchor@muxshape@out\endcsname{%
						\noexpand\muxshape@port[\the\portcount]{1}
					}%
				}{}%
			\fi
		\fi
		\pgfmathaddtocount{\portcount}{1}	
		\repeatpgfmathloop
	}
}

\def\muxshape@port[#1]#2{
	\northeast \pgf@xa=\pgf@x \pgf@ya=\pgf@y
	\southwest \pgf@xb=\pgf@x \pgf@yb=\pgf@y

	\ifnum#2=0
		\ifnum\inverted=0
			\def\chooseports{0}
		\else
			\def\chooseports{1}
		\fi
	\else
		\ifnum\inverted=0
			\def\chooseports{1}
		\else
			\def\chooseports{0}
		\fi
	\fi

	\ifnum\chooseports=0	
		\if\direction\direce
			\pgf@x=\pgf@xb
			\pgf@yc=\pgf@ya \advance\pgf@yc by -\pgf@yb	
			\pgfmathsetlength{\pgf@y}{\pgf@ya-(#1 + 0.5)*(\pgf@yc/\nports)}%
		\fi
		\if\direction\direcw
			\pgf@x=\pgf@xa
			\pgf@yc=\pgf@ya \advance\pgf@yc by -\pgf@yb	
			\pgfmathsetlength{\pgf@y}{\pgf@ya-(#1 + 0.5)*(\pgf@yc/\nports)}%
		\fi
		\if\direction\direcn
			\pgf@y=\pgf@yb
			\pgf@xc=\pgf@xa \advance\pgf@xc by -\pgf@xb	
			\pgfmathsetlength{\pgf@x}{\pgf@xb+(#1 + 0.5)*(\pgf@xc/\nports)}%
		\fi
		\if\direction\direcs
			\pgf@y=\pgf@ya
			\pgf@xc=\pgf@xa \advance\pgf@xc by -\pgf@xb	
			\pgfmathsetlength{\pgf@x}{\pgf@xb+(#1 + 0.5)*(\pgf@xc/\nports)}%
		\fi
	\else	
		\if\direction\direce
			\pgf@x=\pgf@xa
			\pgf@yc=\pgf@ya \advance\pgf@yc by -\pgf@yb	
			\pgfmathsetlength{\pgf@y}{\pgf@ya-0.5\pgf@yc}%
		\fi
		\if\direction\direcw
			\pgf@x=\pgf@xb
			\pgf@yc=\pgf@ya \advance\pgf@yc by -\pgf@yb	
			\pgfmathsetlength{\pgf@y}{\pgf@ya-0.5\pgf@yc}%
		\fi
		\if\direction\direcn
			\pgf@y=\pgf@ya
			\pgf@xc=\pgf@xa \advance\pgf@xc by -\pgf@xb	
			\pgfmathsetlength{\pgf@x}{\pgf@xa-0.5\pgf@xc}%
		\fi
		\if\direction\direcs
			\pgf@y=\pgf@yb
			\pgf@xc=\pgf@xa \advance\pgf@xc by -\pgf@xb	
			\pgfmathsetlength{\pgf@x}{\pgf@xa-0.5\pgf@xc}%
		\fi
	\fi
}

\pgfaddtoshape{muxshape}{
	\anchorlet{c}{center}
	\anchorlet{n}{north}
	\anchorlet{e}{east}
	\anchorlet{s}{south}
	\anchorlet{w}{west}
	\anchorlet{se}{south east}
	\anchorlet{sw}{south west}
	\anchorlet{ne}{north east}
	\anchorlet{nw}{north west}
	\anchorlet{wsw}{west south west}
	\anchorlet{wnw}{west north west}
	\anchorlet{ene}{east north east}
	\anchorlet{ese}{east south east}
	\anchorlet{nnw}{north north west}
	\anchorlet{nne}{north north east}
	\anchorlet{ssw}{south south west}
	\anchorlet{sse}{south south east}
}

\tikzset{
/tikz/muxkeys/.cd,
height/.initial=1,
width/.initial=0.5,
color/.initial=O,
direction/.initial=e,
linestyle/.initial={linestyle, rounded corners = 0},
nports/.initial=3,
inverted/.initial=0,	
smux/.initial=0,
angle/.initial=60,
fillgradient/.initial=O,
/tikz/mux/.code={
\pgfqkeys{/tikz/muxkeys}{#1}%
\tikzset{/tikz/muxkeys/drawer/.expanded=%
\if\pgfkeysvalueof{/tikz/muxkeys/direction}e
	{\pgfkeysvalueof{/tikz/muxkeys/width}}%
	{\pgfkeysvalueof{/tikz/muxkeys/height}}%
	{0}%
	{-90}%
\fi
\if\pgfkeysvalueof{/tikz/muxkeys/direction}w
	{\pgfkeysvalueof{/tikz/muxkeys/width}}%
	{\pgfkeysvalueof{/tikz/muxkeys/height}}%
	{0}%
	{90}%
\fi
\if\pgfkeysvalueof{/tikz/muxkeys/direction}n
	{\pgfkeysvalueof{/tikz/muxkeys/width}}%
	{\pgfkeysvalueof{/tikz/muxkeys/height}}%
	{1}%
	{0}%
\fi
\if\pgfkeysvalueof{/tikz/muxkeys/direction}s
	{\pgfkeysvalueof{/tikz/muxkeys/width}}%
	{\pgfkeysvalueof{/tikz/muxkeys/height}}%
	{1}%
	{180}%
\fi
{\pgfkeysvalueof{/tikz/muxkeys/color}}%
{\pgfkeysvalueof{/tikz/muxkeys/linestyle}}%
{\pgfkeysvalueof{/tikz/muxkeys/smux}}%
{\pgfkeysvalueof{/tikz/muxkeys/angle}}%
{\pgfkeysvalueof{/tikz/muxkeys/fillgradient}}%
}
},
/tikz/muxkeys/drawer/.code n args={9}{%
		\tikzset{
			muxshape,
			#6,
			#5,
			minimum height=
			\ifnum#3>0	
				#1*\NODESIZE
			\else
				#2*\NODESIZE
			\fi
			,minimum width=
			\ifnum#3>0
				#2*\NODESIZE
			\else
				#1*\NODESIZE
			\fi
			,append after command={
					\pgfextra{\let\bdr=\tikzlastnode%
						\node[trapezium, line width = \NODETHICKNESS, minimum height=#1*\NODESIZE, minimum width=#2*\NODESIZE, trapezium stretches=true, rotate=#4, trapezium angle=#8, inner sep=0.001mm] at (\bdr) (trap) {};	
						\ifnum#7=0
							\draw[#9, #5, #6] (trap.bottom left corner) to (trap.top left corner) to (trap.top right corner) to (trap.bottom right corner) to cycle;
						\else
							\draw[#9, #5, #6] (trap.bottom left corner) to[in=#4-90, out=#4+90] (trap.top left corner) to (trap.top right corner) to[in=#4+90,out=#4-90] (trap.bottom right corner) to cycle;
						\fi
					}
				}
		}
	},
}
\newcount\portcount

\pgfdeclareshape{polswitchshape}{
	\savedmacro\nin{
		\edef\nin{\pgfkeysvalueof{/tikz/polswitchkeys/nin}}%
	}
	\savedmacro\nout{
		\edef\nout{\pgfkeysvalueof{/tikz/polswitchkeys/nout}}%
	}
	\savedmacro\direction{
		\edef\direction{\pgfkeysvalueof{/tikz/polswitchkeys/direction}}%
	}
	\inheritsavedanchors[from=basic] 
	
	\inheritanchor[from=basic]{center}
	\inheritanchor[from=basic]{mid}		
	\inheritanchor[from=basic]{base}	
	\inheritanchor[from=basic]{north}
	\inheritanchor[from=basic]{south}		
	\inheritanchor[from=basic]{west}		
	\inheritanchor[from=basic]{mid west}				
	\inheritanchor[from=basic]{base west}		
	\inheritanchor[from=basic]{north west}		
	\inheritanchor[from=basic]{south west}		
	\inheritanchor[from=basic]{east}
	\inheritanchor[from=basic]{mid east}
	\inheritanchor[from=basic]{base east}	
	\inheritanchor[from=basic]{north east}
	\inheritanchor[from=basic]{south east}
	\inheritanchor[from=basic]{west south west}%
	\inheritanchor[from=basic]{west north west}%
	\inheritanchor[from=basic]{east north east}%
	\inheritanchor[from=basic]{east south east}%
	\inheritanchor[from=basic]{north north west}%
	\inheritanchor[from=basic]{north north east}%
	\inheritanchor[from=basic]{south south west}%
	\inheritanchor[from=basic]{south south east}%
	
	\inheritanchorborder[from=basic]	
	\inheritbackgroundpath[from=basic]

    \pgfutil@g@addto@macro\pgf@sh@s@polswitchshape{%
        \pgfmathsetcount{\portcount}{0}
        \pgfmathloop%
        \ifnum\the\portcount<\nin
	        \pgfutil@ifundefined{pgf@anchor@polswitchshape@in\the\portcount}{
		        \expandafter\xdef\csname pgf@anchor@polswitchshape@in\the\portcount\endcsname{%
		            \noexpand\polswitchshape@port[\the\portcount]{0}
		        }%
		    }{}%
	        \ifnum\the\portcount=0
    		    \pgfutil@ifundefined{pgf@anchor@polswitchshape@in}{%
		        \expandafter\xdef\csname pgf@anchor@polswitchshape@in\endcsname{%
		            \noexpand\polswitchshape@port[\the\portcount]{0}
		        }%
		        }{}%
		    \fi
	        \pgfmathaddtocount{\portcount}{1}	
	        \repeatpgfmathloop
        \pgfmathsetcount{\portcount}{0}
        \pgfmathloop%
    	\ifnum\the\portcount<\nout
	        \pgfutil@ifundefined{pgf@anchor@polswitchshape@out\the\portcount}{%
		        \expandafter\xdef\csname pgf@anchor@polswitchshape@out\the\portcount\endcsname{%
		            \noexpand\polswitchshape@port[\the\portcount]{1}
		        }%
		    }{}%
	        \ifnum\the\portcount=0
    		    \pgfutil@ifundefined{pgf@anchor@polswitchshape@out}{%
		        \expandafter\xdef\csname pgf@anchor@polswitchshape@out\endcsname{%
		            \noexpand\polswitchshape@port[\the\portcount]{1}
		        }%
		        }{}%
		    \fi
	        \pgfmathaddtocount{\portcount}{1}	
	        \repeatpgfmathloop
	}
}

\def\polswitchshape@port[#1]#2{
    \northeast \pgf@xa=\pgf@x \pgf@ya=\pgf@y
    \southwest \pgf@xb=\pgf@x \pgf@yb=\pgf@y
    
    \ifnum#2=0	
	    \if\direction\direce	
	    	\pgf@x=\pgf@xb
		    \pgf@yc=\pgf@ya \advance\pgf@yc by -\pgf@yb	
		    \pgfmathsetlength{\pgf@y}{\pgf@ya-(#1 + 0.5)*(\pgf@yc/\nin)}%
	    \fi
	    \if\direction\direcw
	    	\pgf@x=\pgf@xa
		    \pgf@yc=\pgf@ya \advance\pgf@yc by -\pgf@yb	
		    \pgfmathsetlength{\pgf@y}{\pgf@ya-(#1 + 0.5)*(\pgf@yc/\nin)}%
	    \fi
	    \if\direction\direcn
	    	\pgf@y=\pgf@yb
		    \pgf@xc=\pgf@xa \advance\pgf@xc by -\pgf@xb	
		    \pgfmathsetlength{\pgf@x}{\pgf@xb+(#1 + 0.5)*(\pgf@xc/\nin)}%
	    \fi
	    \if\direction\direcs
	    	\pgf@y=\pgf@ya
		    \pgf@xc=\pgf@xa \advance\pgf@xc by -\pgf@xb	
		    \pgfmathsetlength{\pgf@x}{\pgf@xb+(#1 + 0.5)*(\pgf@xc/\nin)}%
	    \fi
	\else	
	    \if\direction\direce	
	    	\pgf@x=\pgf@xa
		    \pgf@yc=\pgf@ya \advance\pgf@yc by -\pgf@yb	
		    \pgfmathsetlength{\pgf@y}{\pgf@ya-(#1 + 0.5)*(\pgf@yc/\nout)}%
	    \fi
	    \if\direction\direcw
	    	\pgf@x=\pgf@xb
		    \pgf@yc=\pgf@ya \advance\pgf@yc by -\pgf@yb	
		    \pgfmathsetlength{\pgf@y}{\pgf@ya-(#1 + 0.5)*(\pgf@yc/\nout)}%
	    \fi
	    \if\direction\direcn
	    	\pgf@y=\pgf@ya
		    \pgf@xc=\pgf@xa \advance\pgf@xc by -\pgf@xb	
		    \pgfmathsetlength{\pgf@x}{\pgf@xb+(#1 + 0.5)*(\pgf@xc/\nout)}%
	    \fi
	    \if\direction\direcs
	    	\pgf@y=\pgf@yb
		    \pgf@xc=\pgf@xa \advance\pgf@xc by -\pgf@xb	
		    \pgfmathsetlength{\pgf@x}{\pgf@xb+(#1 + 0.5)*(\pgf@xc/\nout)}%
	    \fi
	\fi
}

\pgfaddtoshape{polswitchshape}{
	\anchorlet{c}{center}		
	\anchorlet{n}{north}
	\anchorlet{e}{east}
	\anchorlet{s}{south}
	\anchorlet{w}{west}			
	\anchorlet{se}{south east}
	\anchorlet{sw}{south west}
	\anchorlet{ne}{north east}
	\anchorlet{nw}{north west}
	\anchorlet{wsw}{west south west}
	\anchorlet{wnw}{west north west}
	\anchorlet{ene}{east north east}
	\anchorlet{ese}{east south east}
	\anchorlet{nnw}{north north west}
	\anchorlet{nne}{north north east}
	\anchorlet{ssw}{south south west}
	\anchorlet{sse}{south south east}
}

\tikzset{
	/tikz/polswitchkeys/.cd,
	size/.initial=1,
	color/.initial=O,
	direction/.initial=e,
	linestyle/.initial={linestyle, inner sep=0.5mm},
	nin/.initial=1,	
	nout/.initial=1, 
	/tikz/polswitch/.code={
		\pgfqkeys{/tikz/polswitchkeys}{#1}%
		\tikzset{/tikz/polswitchkeys/drawer/.expanded=%
			{\pgfkeysvalueof{/tikz/polswitchkeys/size}}%
			{\pgfkeysvalueof{/tikz/polswitchkeys/color}}%
			{\pgfkeysvalueof{/tikz/polswitchkeys/linestyle}}%
			{\pgfkeysvalueof{/tikz/polswitchkeys/nout}}%
			\if\pgfkeysvalueof{/tikz/polswitchkeys/direction}e
				{0}%
			\fi
			\if\pgfkeysvalueof{/tikz/polswitchkeys/direction}w
				{0}%
			\fi
			\if\pgfkeysvalueof{/tikz/polswitchkeys/direction}n
				{1}%
			\fi
			\if\pgfkeysvalueof{/tikz/polswitchkeys/direction}s
				{1}%
			\fi
			{\pgfkeysvalueof{/tikz/polswitchkeys/direction}}%
		}
	},
	/tikz/polswitchkeys/drawer/.code n args={6}{%
		\tikzset{
			polswitchshape,
			draw,
			minimum height = #1*\NODESIZE,
			minimum width = #1*\NODESIZE,
			#2,
			#3,
			append after command={
				\pgfextra{\let\bdr=\tikzlastnode%
						\node[coordinate] at ($(\bdr.in)!0.25!(\bdr.out)$) (circlein){};
						\node[coordinate] at ($(\bdr.in)!0.75!(\bdr.out)$) (circleout){};

						\draw[---, #2, #3, fill] (\bdr.in) to (circlein) circle (0.05);
						\draw[---, #2, #3, fill] (\bdr.out) to (circleout) circle (0.05);

						\node[coordinate] at ($(\bdr.in)!0.6!(\bdr.out)$) (circlemiddle){};
						\ifnum#5>0
							\node[coordinate] at ($(circlemiddle)!0.5!(circlemiddle -| \bdr.e)$) (circletopcor){};
							\node[coordinate] at ($(circlemiddle)!0.5!(circlemiddle -| \bdr.w)$) (circlebotcor){};
						\else
							\node[coordinate] at ($(circlemiddle)!0.5!(circlemiddle |- \bdr.n)$) (circletopcor){};
							\node[coordinate] at ($(circlemiddle)!0.5!(circlemiddle |- \bdr.s)$) (circlebotcor){};
						\fi

						\node[draw, circle, #2, #3, minimum size=0.3*\FNODESIZE] at (circletopcor) (circletop){};
						\node[draw, circle, #2, #3, minimum size=0.3*\FNODESIZE] at (circlebotcor) (circlebot){};
						\draw[-->, #2, #3] (circletop.south) to (circletop.north){};
						\draw[-->, #2, #3] (circlebot.west) to (circlebot.east){};

						\if#6e
							\draw[-->, #2, #3] ([xshift=-0.05*\FNODESIZE]circletop.south west) to [out=-120, in=120] ([xshift=-0.05*\FNODESIZE]circlebot.north west){};
						\fi
						\if#6w
							\draw[-->, #2, #3] ([xshift=0.05*\FNODESIZE]circletop.south east) to [out=-60, in=60] ([xshift=0.05*\FNODESIZE]circlebot.north east){};
						\fi
						\if#6s
							\draw[-->, #2, #3] ([yshift=0.05*\FNODESIZE]circletop.north west) to [out=150, in=30] ([yshift=0.05*\FNODESIZE]circlebot.north east){};
						\fi
						\if#6n
							\draw[-->, #2, #3] ([yshift=-0.05*\FNODESIZE]circletop.south west) to [out=-150, in=-30] ([yshift=-0.05*\FNODESIZE]circlebot.south east){};
						\fi

				}
			}
		}
	},
}
\pgfdeclareshape{pdshape}{
	\savedmacro\direction{
		\edef\direction{\pgfkeysvalueof{/tikz/pdkeys/direction}}%
	}
	\saveddimen\minwidth{
		\pgfmathsetlength\pgf@x{\pgfshapeminwidth}%
	}
	\saveddimen\minheight{
		\pgfmathsetlength\pgf@x{\pgfshapeminheight}%
	}
	\inheritsavedanchors[from=basic]

	\inheritanchor[from=basic]{center}
	\inheritanchor[from=basic]{mid}
	\inheritanchor[from=basic]{base}
	\inheritanchor[from=basic]{north}
	\inheritanchor[from=basic]{south}
	\inheritanchor[from=basic]{west}
	\inheritanchor[from=basic]{mid west}
	\inheritanchor[from=basic]{base west}
	\inheritanchor[from=basic]{north west}
	\inheritanchor[from=basic]{south west}
	\inheritanchor[from=basic]{east}
	\inheritanchor[from=basic]{mid east}
	\inheritanchor[from=basic]{base east}
	\inheritanchor[from=basic]{north east}
	\inheritanchor[from=basic]{south east}
	\inheritanchor[from=basic]{west south west}%
	\inheritanchor[from=basic]{west north west}%
	\inheritanchor[from=basic]{east north east}%
	\inheritanchor[from=basic]{east south east}%
	\inheritanchor[from=basic]{north north west}%
	\inheritanchor[from=basic]{north north east}%
	\inheritanchor[from=basic]{south south west}%
	\inheritanchor[from=basic]{south south east}%

	\inheritanchorborder[from=basic]
	\inheritbackgroundpath[from=basic]

	\pgfutil@g@addto@macro\pgf@sh@s@pdshape{%
		\pgfutil@ifundefined{pgf@anchor@pdshape@in0}{
			\expandafter\xdef\csname pgf@anchor@pdshape@in0\endcsname{%
				\noexpand\pdshape@port{0}
			}%
		}{}%
		\pgfutil@ifundefined{pgf@anchor@pdshape@in}{
			\expandafter\xdef\csname pgf@anchor@pdshape@in\endcsname{%
				\noexpand\pdshape@port{0}
			}%
		}{}%
		\pgfutil@ifundefined{pgf@anchor@pdshape@out0}{
			\expandafter\xdef\csname pgf@anchor@pdshape@out0\endcsname{%
				\noexpand\pdshape@port{1}
			}%
		}{}%
		\pgfutil@ifundefined{pgf@anchor@pdshape@out}{
			\expandafter\xdef\csname pgf@anchor@pdshape@out\endcsname{%
				\noexpand\pdshape@port{1}
			}%
		}{}%
	}
}

\def\pdshape@port#1{
	\northeast	

	\ifnum#1=0	
		\if\direction\direce
			\pgf@x=-\pgf@x
			\pgf@ya= \pgf@y
			\pgfmathsetlength{\pgf@y}{\pgf@ya-0.5*\minheight}%
		\fi
		\if\direction\direcw
			\pgf@x=\pgf@x
			\pgf@ya= \pgf@y
			\pgfmathsetlength{\pgf@y}{\pgf@ya-0.5*\minheight}%
		\fi
		\if\direction\direcn
			\pgf@y=-\pgf@y
			\pgf@xa=\pgf@x
			\pgfmathsetlength{\pgf@x}{\pgf@xa-0.5*\minwidth}%
		\fi
		\if\direction\direcs
			\pgf@y=\pgf@y
			\pgf@xa= \pgf@x
			\pgfmathsetlength{\pgf@x}{\pgf@xa-0.5*\minwidth}%
		\fi
	\else	
		\if\direction\direce
			\pgf@x=\pgf@x
			\pgf@ya= \pgf@y
			\pgfmathsetlength{\pgf@y}{\pgf@ya-0.5*\minheight}%
		\fi
		\if\direction\direcw
			\pgf@x=-\pgf@x
			\pgf@ya= \pgf@y
			\pgfmathsetlength{\pgf@y}{\pgf@ya-0.5*\minheight}%
		\fi
		\if\direction\direcn
			\pgf@y=\pgf@y
			\pgf@xa= \pgf@x
			\pgfmathsetlength{\pgf@x}{\pgf@xa-0.5*\minwidth}%
		\fi
		\if\direction\direcs
			\pgf@y=-\pgf@y
			\pgf@xa= \pgf@x
			\pgfmathsetlength{\pgf@x}{\pgf@xa-0.5*\minwidth}%
		\fi
	\fi
}

\pgfaddtoshape{pdshape}{
	\anchorlet{c}{center}
	\anchorlet{n}{north}
	\anchorlet{e}{east}
	\anchorlet{s}{south}
	\anchorlet{w}{west}
	\anchorlet{se}{south east}
	\anchorlet{sw}{south west}
	\anchorlet{ne}{north east}
	\anchorlet{nw}{north west}
	\anchorlet{wsw}{west south west}
	\anchorlet{wnw}{west north west}
	\anchorlet{ene}{east north east}
	\anchorlet{ese}{east south east}
	\anchorlet{nnw}{north north west}
	\anchorlet{nne}{north north east}
	\anchorlet{ssw}{south south west}
	\anchorlet{sse}{south south east}
}

\tikzset{
/tikz/pdkeys/.cd,
size/.initial=0.5,
color/.initial=EO,
direction/.initial=e,
linestyle/.initial={linestyle},
fillgradient/.initial=O,
/tikz/pd/.code={
		\pgfqkeys{/tikz/pdkeys}{#1}%
		\tikzset{/tikz/pdkeys/drawer/.expanded=%
				{\pgfkeysvalueof{/tikz/pdkeys/direction}}%
				{\pgfkeysvalueof{/tikz/pdkeys/size}}%
				{\pgfkeysvalueof{/tikz/pdkeys/color}}%
				{\pgfkeysvalueof{/tikz/pdkeys/linestyle}}%
				{\pgfkeysvalueof{/tikz/pdkeys/fillgradient}}%
		}
	},
/tikz/pdkeys/drawer/.code n args={5}{%
\tikzset{
pdshape,
minimum height=#2*\NODESIZE,
minimum width=#2*\NODESIZE,
#3,
#4,
draw,
append after command={
\pgfextra{\let\bdr=\tikzlastnode%
\node[#5, fit=(\bdr.nw)(\bdr.se)] (boxgradient){};
\draw[---,#3] ($(\bdr.s)!.1!(\bdr.n)$) to ($(\bdr.s)!.9!(\bdr.n)$);
\fill[#3] ({$(\bdr.s)!.3!(\bdr.n)$} -| {$(\bdr.w)!.3!(\bdr.e)$}) to ($(\bdr.s)!.7!(\bdr.n)$) to ({$(\bdr.s)!.3!(\bdr.n)$} -| {$(\bdr.w)!.7!(\bdr.e)$}) to cycle;
\draw[---,#3] ({$(\bdr.s)!.7!(\bdr.n)$} -| {$(\bdr.w)!.35!(\bdr.e)$}) to ({$(\bdr.s)!.7!(\bdr.n)$} -| {$(\bdr.w)!.65!(\bdr.e)$});
}
}
}
},
}
\pgfdeclareshape{pbsshape}{
	\savedmacro\direction{
		\edef\direction{\pgfkeysvalueof{/tikz/pbskeys/direction}}%
	}
	\saveddimen\minwidth{
		\pgfmathsetlength\pgf@x{\pgfshapeminwidth}%
	}
	\saveddimen\minheight{
		\pgfmathsetlength\pgf@x{\pgfshapeminheight}%
	}
	\savedmacro\nport{
		\edef\nport{\pgfkeysvalueof{/tikz/pbskeys/nport}}%
	}
	\inheritsavedanchors[from=basic]

	\inheritanchor[from=basic]{center}
	\inheritanchor[from=basic]{mid}
	\inheritanchor[from=basic]{base}
	\inheritanchor[from=basic]{north}
	\inheritanchor[from=basic]{south}
	\inheritanchor[from=basic]{west}
	\inheritanchor[from=basic]{mid west}
	\inheritanchor[from=basic]{base west}
	\inheritanchor[from=basic]{north west}
	\inheritanchor[from=basic]{south west}
	\inheritanchor[from=basic]{east}
	\inheritanchor[from=basic]{mid east}
	\inheritanchor[from=basic]{base east}
	\inheritanchor[from=basic]{north east}
	\inheritanchor[from=basic]{south east}
	\inheritanchor[from=basic]{west south west}%
	\inheritanchor[from=basic]{west north west}%
	\inheritanchor[from=basic]{east north east}%
	\inheritanchor[from=basic]{east south east}%
	\inheritanchor[from=basic]{north north west}%
	\inheritanchor[from=basic]{north north east}%
	\inheritanchor[from=basic]{south south west}%
	\inheritanchor[from=basic]{south south east}%

	\inheritanchorborder[from=basic]
	\inheritbackgroundpath[from=basic]

	\pgfutil@g@addto@macro\pgf@sh@s@pbsshape{%
		\pgfutil@ifundefined{pgf@anchor@pbsshape@in0}{
			\expandafter\xdef\csname pgf@anchor@pbsshape@in0\endcsname{%
				\noexpand\pbsshape@port[0]{0}
			}%
		}{}%
		\pgfutil@ifundefined{pgf@anchor@pbsshape@in1}{
			\expandafter\xdef\csname pgf@anchor@pbsshape@in1\endcsname{%
				\noexpand\pbsshape@port[1]{0}
			}%
		}{}%
		\pgfutil@ifundefined{pgf@anchor@pbsshape@in}{
			\expandafter\xdef\csname pgf@anchor@pbsshape@in\endcsname{%
				\noexpand\pbsshape@port[0]{0}
			}%
		}{}%
		\pgfutil@ifundefined{pgf@anchor@pbsshape@out0}{
			\expandafter\xdef\csname pgf@anchor@pbsshape@out0\endcsname{%
				\noexpand\pbsshape@port[0]{1}
			}%
		}{}%
		\pgfutil@ifundefined{pgf@anchor@pbsshape@out1}{
			\expandafter\xdef\csname pgf@anchor@pbsshape@out1\endcsname{%
				\noexpand\pbsshape@port[1]{1}
			}%
		}{}%
		\pgfutil@ifundefined{pgf@anchor@pbsshape@out}{
			\expandafter\xdef\csname pgf@anchor@pbsshape@out\endcsname{%
				\noexpand\pbsshape@port[0]{1}
			}%
		}{}%
		\pgfmathsetcount{\portcount}{0}
		\pgfmathloop%
		\ifnum\the\portcount<\nport
		\pgfutil@ifundefined{pgf@anchor@pbsshape@p\the\portcount}{
			\expandafter\xdef\csname pgf@anchor@pbsshape@p\the\portcount\endcsname{%
				\noexpand\pbsshape@port[\the\portcount]{2}
			}%
		}{}%
		\pgfmathaddtocount{\portcount}{1}	
		\repeatpgfmathloop
	}
}

\def\pbsshape@port[#1]#2{
	\northeast	

	\ifnum#2=0	
		\if\direction\direce
			\ifnum#1=0
				\pgf@x=-\pgf@x
				\pgf@ya= \pgf@y
				\pgfmathsetlength{\pgf@y}{\pgf@ya-0.5*\minheight}%
			\else
				\pgf@x=0\pgf@x
				\pgf@ya= \pgf@y
				\pgfmathsetlength{\pgf@y}{\pgf@ya-\minheight}%
			\fi
		\fi
		\if\direction\direcw
			\ifnum#1=0
				\pgf@x=\pgf@x
				\pgf@ya= \pgf@y
				\pgfmathsetlength{\pgf@y}{\pgf@ya-0.5*\minheight}%
			\else
				\pgf@x=0\pgf@x
				\pgf@ya= \pgf@y
				\pgfmathsetlength{\pgf@y}{\pgf@ya}%
			\fi
		\fi
		\if\direction\direcn
			\ifnum#1=0
				\pgf@y=-\pgf@y
				\pgf@xa=\pgf@x
				\pgfmathsetlength{\pgf@x}{\pgf@xa-0.5*\minwidth}%
			\else
				\pgf@y=0\pgf@y
				\pgf@xa=\pgf@x
				\pgfmathsetlength{\pgf@x}{\pgf@xa-0*\minwidth}%
			\fi
		\fi
		\if\direction\direcs
			\ifnum#1=0
				\pgf@y=\pgf@y
				\pgf@xa= \pgf@x
				\pgfmathsetlength{\pgf@x}{\pgf@xa-0.5*\minwidth}%
			\else
				\pgf@y=0\pgf@y
				\pgf@xa= \pgf@x
				\pgfmathsetlength{\pgf@x}{\pgf@xa-1*\minwidth}%
			\fi
		\fi
	\else	
		\if\direction\direce
			\ifnum#1=0
				\pgf@x=\pgf@x
				\pgf@ya= \pgf@y
				\pgfmathsetlength{\pgf@y}{\pgf@ya-0.5*\minheight}%
			\else
				\pgf@x=0\pgf@x
				\pgf@ya= \pgf@y
				\pgfmathsetlength{\pgf@y}{\pgf@ya}%
			\fi
		\fi
		\if\direction\direcw
			\ifnum#1=0
				\pgf@x=-\pgf@x
				\pgf@ya= \pgf@y
				\pgfmathsetlength{\pgf@y}{\pgf@ya-0.5*\minheight}%
			\else
				\pgf@x=0\pgf@x
				\pgf@ya= \pgf@y
				\pgfmathsetlength{\pgf@y}{\pgf@ya-\minheight}%
			\fi
		\fi
		\if\direction\direcn
			\ifnum#1=0
				\pgf@y=\pgf@y
				\pgf@xa= \pgf@x
				\pgfmathsetlength{\pgf@x}{\pgf@xa-0.5*\minwidth}%
			\else
				\pgf@y=0\pgf@y
				\pgf@xa= \pgf@x
				\pgfmathsetlength{\pgf@x}{\pgf@xa-1*\minwidth}%
			\fi
		\fi
		\if\direction\direcs
			\ifnum#1=0
				\pgf@y=-\pgf@y
				\pgf@xa= \pgf@x
				\pgfmathsetlength{\pgf@x}{\pgf@xa-0.5*\minwidth}%
			\else
				\pgf@y=0\pgf@y
				\pgf@xa= \pgf@x
				\pgfmathsetlength{\pgf@x}{\pgf@xa-0*\minwidth}%
			\fi
		\fi
	\fi

	\ifnum#2=2	%
		\northeast \pgf@xa=\pgf@x \pgf@ya=\pgf@y
		\southwest \pgf@xb=\pgf@x \pgf@yb=\pgf@y

		\pgf@xc=\pgf@xa \advance\pgf@xc by -\pgf@xb	
		\pgfmathsetlength{\pgf@x}{\pgf@xa-(#1 + 0.5)*(\pgf@xc/\nport)}%
		\pgf@yc=\pgf@ya \advance\pgf@yc by -\pgf@yb	
		\pgfmathsetlength{\pgf@y}{\pgf@ya-(#1 + 0.5)*(\pgf@yc/\nport)}%

		\if\direction\direce
			\pgf@x=-\pgf@x
		\fi
		\if\direction\direcw
			\pgf@x=-\pgf@x
		\fi
	\fi
}

\pgfaddtoshape{pbsshape}{
	\anchorlet{c}{center}
	\anchorlet{n}{north}
	\anchorlet{e}{east}
	\anchorlet{s}{south}
	\anchorlet{w}{west}
	\anchorlet{se}{south east}
	\anchorlet{sw}{south west}
	\anchorlet{ne}{north east}
	\anchorlet{nw}{north west}
	\anchorlet{wsw}{west south west}
	\anchorlet{wnw}{west north west}
	\anchorlet{ene}{east north east}
	\anchorlet{ese}{east south east}
	\anchorlet{nnw}{north north west}
	\anchorlet{nne}{north north east}
	\anchorlet{ssw}{south south west}
	\anchorlet{sse}{south south east}
}

\pgfmathsetmacro{\WSSSINEHEIGHT}{0.06}

\tikzset{
	/tikz/pbskeys/.cd,
	size/.initial=0.5,
	color/.initial=O,
	direction/.initial=e,
	linestyle/.initial={linestyle},
	nport/.initial=1,
	fillgradient/.initial=O,
	/tikz/pbs/.code={
			\pgfqkeys{/tikz/pbskeys}{#1}%
			\tikzset{/tikz/pbskeys/drawer/.expanded=%
					{\pgfkeysvalueof{/tikz/pbskeys/direction}}%
					{\pgfkeysvalueof{/tikz/pbskeys/size}}%
					{\pgfkeysvalueof{/tikz/pbskeys/color}}%
					{\pgfkeysvalueof{/tikz/pbskeys/linestyle}}%
					{\pgfkeysvalueof{/tikz/pbskeys/fillgradient}}%
			}
		},
	/tikz/pbskeys/drawer/.code n args={5}{%
			\tikzset{
				pbsshape,
				minimum height=#2*\NODESIZE,
				minimum width=#2*\NODESIZE,
				#3,
				#4,
				draw,
				append after command={
						\pgfextra{\let\bdr=\tikzlastnode%
							\node[#5, fit=(\bdr.nw)(\bdr.se)] (boxgradient){};

							\if#1e
								\draw[#3, ---] ($(\bdr.nw)!.01!(\bdr.se)$) to ($(\bdr.se)!.01!(\bdr.nw)$);
							\fi
							\if#1w
								\draw[#3, ---] ($(\bdr.nw)!.01!(\bdr.se)$) to ($(\bdr.se)!.01!(\bdr.nw)$);
							\fi
							\if#1n
								\draw[#3, ---] ($(\bdr.ne)!.01!(\bdr.sw)$) to ($(\bdr.sw)!.01!(\bdr.ne)$);
							\fi
							\if#1s
								\draw[#3, ---] ($(\bdr.ne)!.01!(\bdr.sw)$) to ($(\bdr.sw)!.01!(\bdr.ne)$);
							\fi

						}
					}
			}
		},
}

\ifdefined\fontchoice
\else
	\def\fontchoice{times} 
\fi

\usepackage{pdftexcmds}

\ifnum\pdf@strcmp{\fontchoice}{firasans}=0 %
	\usepackage[book]{FiraSans} 
	\usepackage[T1]{fontenc}
	
	\tikzset{every picture/.style={/utils/exec={\sffamily}}}
\fi

\ifnum\pdf@strcmp{\fontchoice}{times}=0 %
	\usepackage{times}
\fi

\ifnum\pdf@strcmp{\fontchoice}{timesnewroman}=0 %
	\usepackage{mathptmx}
	\usepackage[T1]{fontenc}
\fi

\ifnum\pdf@strcmp{\fontchoice}{helvetica}=0 %
	\usepackage[scaled]{helvet}
	
	\usepackage[T1]{fontenc}
\fi

\makeatother
\pgfplotscreateplotcyclelist{foo}{
        {C1,mark=*},
        {C2,mark=square*},
        {C3, mark = triangle*},
        {C4,mark=+},
        {C5, mark = diamond*},
        {C6, mark = x}
      }
\begin{document}

\title{Adaptive Reconciliation for Experimental Continuous-Variable Quantum Key Distribution Over a Turbulent Free-Space Optical Channel\vspace{-5mm}}


\author{Kadir G\" um\" u\c s\authormark{(1,*)}, Jo\~ ao dos Reis Fraz\~ ao,\authormark{(1)}, Vincent van Vliet\authormark{(1)}, Sjoerd van der Heide\authormark{(1)}, \\ Menno van den Hout\authormark{(1)}, Aaron Albores-Mejia\authormark{(1,2)}, Thomas Bradley\authormark{(1)}, and Chigo Okonkwo\authormark{(1,2)}}

\address{$^{(1)}$High-capacity Optical Transmission Laboratory, Eindhoven University of Technology, The Netherlands \\ 
\textit{$^{(2)}$} CUbIQ Technologies, De Groene Loper 5, Eindhoven, The Netherlands }

\email{\authormark{*}k.gumus@tue.nl\vspace{-5mm}} 

\begin{abstract}
We experimentally demonstrate adaptive reconciliation for continuous-variable quantum key distribution over a turbulent free-space optical channel. Additionally, we propose a method for optimising the reconciliation efficiency, increasing secret key rates by up to 8.1\%. 
\end{abstract}
\vspace{1mm}
\section{Introduction}
\vspace{-2mm}
\begin{figure}[b!]
    \centering
    \includegraphics{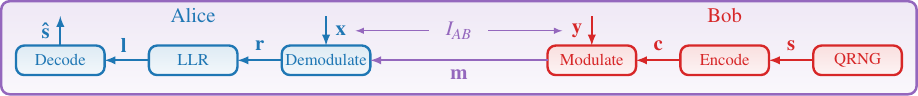}
    \caption{An overview of the reverse multi-dimensional reconciliation protocol for CV-QKD.}
    \label{fig:Reconciliation}
\end{figure}
In recent years, many exciting advancements have been made in the field of quantum computing, with expected improvements in the number of qubits\cite{gyongyosi2019survey}. Although these advancements will certainly carry great benefits, they also raise concerns about data security, as these quantum technologies may break current data security protocols. As a result, attention has shifted to protocols that remain unbreakable, even in a post-quantum world. One such protocol is continuous-variable quantum key distribution (CV-QKD), a method for securely sharing secret keys between two communicating parties, Alice and Bob, without a potential eavesdropper (Eve) recovering keys \cite{Laudenbach_2018}.
\\\indent One of the channels considered for CV-QKD is the free-space optical (FSO) channel, as this would allow for wireless key sharing. The instability of the FSO channel caused by atmospheric turbulence is less of an issue for CV-QKD compared to classical communications, as the encoding occurs after transmission. Despite this, CV-QKD implementation for an FSO channel still poses a challenge, with different considerations and trade-offs to be made when compared to the optical fibre channel. Several experimental CV-QKD transmissions over FSO have already been demonstrated \cite{RANI2023100162,PhysRevA.100.012325,Wang:21}, however, the reconciliation is rarely analysed, often kept as a footnote.
\\\indent In this paper, we demonstrate adaptive reconciliation for CV-QKD over an FSO channel with differing turbulence strengths. We show that the modulation format significantly impacts the reconciliation performance, and by choosing a high-dimensional reconciliation protocol, secret key rates (SKRs) can be increased by 122\%. Finally, we propose a method for optimising the reconciliation efficiency, increasing SKR by an additional 8.1\%. 
\vspace{-2mm}
\section{Reverse Reconciliation}
\vspace{-2mm}
\indent Reconciliation is a part of the CV-QKD protocol where Alice and Bob share bit strings which is employed for generating the secret keys. In this paper we will consider only reverse multi-dimensional reconciliation, as direct reconciliation is affected by the 3dB limit \cite{Laudenbach_2018}, and slice reconciliation performs worse than the multi-dimensional protocol in the regime we operate at\cite{Leverrier_2008}. An overview of reverse multi-dimensional reconciliation as described in \cite{Leverrier_2008} is given in Fig. \ref{fig:Reconciliation}. At the start of the reconciliation, Alice and Bob have the transmitted and measured quantum states $\mathbf{x}$ and $\mathbf{y}$, respectively. Bob generates a string of bits $\mathbf{s}$ using a quantum random number generator (QRNG) and encodes these bits using an error correction code with code rate $R$ creating a codeword $\mathbf{c}$, which is modulated using $\mathbf{y}$ to obfuscate the values of the bits. Bob transmits the modulated message $\mathbf{m}$ to Alice over the classical channel, where it is demodulated before calculation of the log-likelihood ratios (LLR). These LLRs are employed in the decoder to get $\hat{\mathbf{s}}$, which is an estimate of $\mathbf{s}$. After decoding, Alice and Bob compare whether $\mathbf{s}$ and $\hat{\mathbf{s}}$ are the same with, for example, a hashing function. If the decoding has failed, a frame error has occurred and the frame is discarded, otherwise, the frame will be used during privacy amplification for key generation. 
\\\indent The SKR depends on the performance of the error correction codes used during reconciliation and is given by $\text{SKR} = (1-\text{FER})(\beta I_{AB} - \chi_{BE})$, where FER is the frame error rate, $I_{AB}$ is the mutual information between $\mathbf{x}$ and $\mathbf{y}$, $\beta = \frac{R}{I_{AB}}$ is the reconciliation efficiency, and $\chi_{BE}$ is the Holevo information (Eve's information)\cite{Laudenbach_2018}. Secret key exchange is possible when the SKR is positive, i.e., $\beta$ is close to 1. There is a trade-off between $\beta$ and the FER, as the FER increases as $\beta$ increases, so good performing error correction is vital for providing high key rates.
\vspace{-2mm}
\section{Experimental Set-up}
\vspace{-2mm}
\begin{figure}[t!]
    \centering
    \resizebox{0.98\linewidth}{!}{\includegraphics{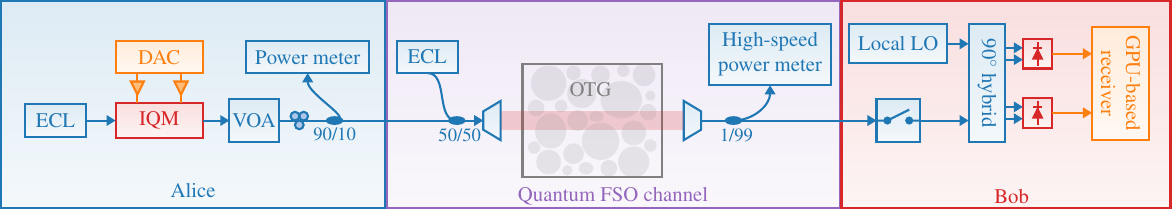}}
    \caption{The CV-QKD set-up for transmission over an FSO channel with an optical turbulence generator.}
    \label{fig:set-up}
\end{figure}
\indent The experimental set-up for the CV-QKD transmission is shown in Fig. \ref{fig:set-up}. Alice employs a \SI{<100}{kHz}  linewidth external cavity laser (ECL) at 1550~nm, a digital-to-analog (DAC) converter, and an optical modulator (IQM) to modulate probabilistically-shaped 256QAM (PS-256QAM) signals \cite{Denys_2021} at a symbol rate of 250 Mbaud. With a variable optical attenuator (VOA) and a power meter, the power of the signal is attenuated to an average of 7.44 shot noise units (SNU) (-69.2~dBm). The attenuated 1550~nm signal is combined with a second tone at 1528~nm and converted to free space using a collimator. The light traverses an optical turbulence generator (OTG) \cite{OTGthesis} which can mimic varying turbulence strengths. Afterwards, the light is coupled back to fiber using a collimator and a portion is split to a high-speed power meter for turbulence characterisation by fitting a combined log-normal pointing jitter distribution according to \cite{1994OptEn..33.3748K}. We measured four different turbulence strengths generated by the OTG, with scintillation index $\sigma_I = $ 0.001, 0.009, 0.01, 0.013 and pointing jitter $\beta_{jitter}$ = 123.8, 8.6, 3.0, 1.6 respectively, all classified as weak fluctuations\cite{1994OptEn..33.3748K}.
\\\indent The remaining 99\% of light is directed to Bob's side. A local ECL is used as a local local oscillator (LLO) for the 90$^{\circ}$ optical hybrid, and the outputs are digitised. Digital signal processing is used for calibration and recovery of the quantum signal \cite{Sjoerd2023}. Parameter estimation, taking into account finite-size effects\cite{Jouguet_2012_fs}, is performed on each CV-QKD block and $I_{AB}$, the excess noise $\xi_{Bob}$, and $\chi_{BE}$ are estimated. Other relevant parameters for the system are a clearance of 10~dB, quantum efficiency $\eta$ of 40\%, CV-QKD block size of $6.8\cdot10^6$, average $\xi_{Bob}$ of 0.0045~SNU, and an average transmittance $T$ depending on the turbulence strength, ranging from 0.35 to 0.41.
\\\indent For the reverse reconciliation protocol, we use high-dimensional reconciliation, which is multi-dimensional reconciliation with dimensionality $d > 8$, with $d = 128$ as described in \cite{Leverrier_2008}. Although normally multi-dimensional reconciliation with $d = 8$ is chosen as it is less complex \cite{Leverrier_2008}, we opted for higher $d$, as using PS-256QAM for the modulation of the quantum states significantly reduces the performance of the error correction. The unequal power of the transmitted quantum states makes it so that the virtual channel created during reconciliation does not exactly resemble the binary input additive white Gaussian noise (BI-AWGN) channel which the error correction codes are designed for\cite{Jouguet_2011}. As $d \to \infty$, the virtual channel converges to the BI-AWGN as the power of the transmitted quantum states gets averaged out, therefore a higher $d$ improves the performance of the reconciliation.
\\\indent We use a $R = 0.2$ expanded type-based protograph low-density parity check (TBP-LDPC) code \cite{gumucs2021low} punctured to $R \approx 0.3$, the average $I_{AB}$ of the system, for error correction. This code was chosen because it operates close to capacity, even after significantly changing the rate of the code after puncturing. To adapt the rate of the code during the reconciliation we use the \textit{sp}-protocol described in \cite{wang2017efficient}. We choose a blocklength $N \approx10^5$ with a maximum of 500 decoding iterations. We randomly sample the parity check matrix according to the protograph, but remove all short cycles within the graph to ensure good error correction performance. 
\\\indent One additional improvement is on how to choose $\beta$. Normally, $\beta$ is determined at a fixed value in order to optimise the average SKR of the system over all CV-QKD blocks. When $I_{AB}$ changes, $R$ is changed such that $\beta$ stays consistent, which is a valid approach for CV-QKD transmission over fibre, as it tends to be a stable channel. However, in an FSO channel atmospheric turbulence causes additional time-dependent instabilities. As a result, the optimal $\beta$ for each CV-QKD block changes, and it would make more sense to adaptively change $\beta$ to optimise the SKR of each block. This adds no extra complexity to the system, as it can be implemented by using a $\beta$-FER look-up table during the parameter estimation phase, and picking the $\beta$-FER pair which maximises the SKR. 
\vspace{-2mm}
\section{Results}
\begin{figure}[t!]
\begin{subfigure}{0.33\textwidth}
    \includegraphics{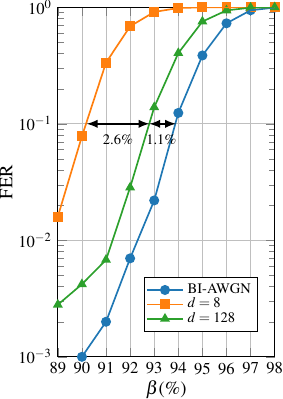}
\end{subfigure}
\begin{subfigure}{0.33\textwidth}
    \includegraphics{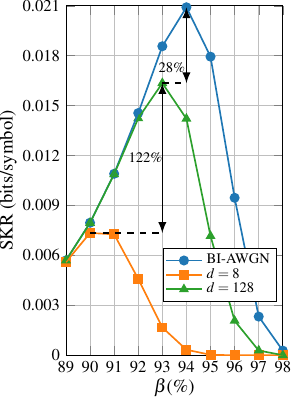}
\end{subfigure}
\begin{subfigure}{0.33\textwidth}
    \includegraphics{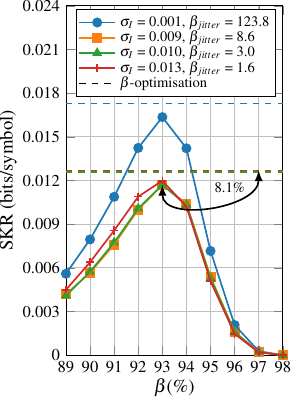}
\end{subfigure}
    \caption{Left: The FER of the $R=\frac{1}{5}$ expanded TBP-LDPC code when punctured to $R = 0.3$ for different $d$ compared to the BI-AWGN channel. Middle: The SKR of the $R=\frac{1}{5}$ expanded TBP-LDPC code when punctured to $R = 0.3$ for different $d$ compared to the BI-AWGN channel for the FSO channel with $\sigma_I$ = 0.001 and $\beta_{jitter}$ = 123.8. Right: The SKR for our CV-QKD set-up for different $\beta$ compared to $\beta$-optimisation (dashed line) for different turbulence settings.}
    \label{FER_SKR}
\end{figure}

\vspace{-2mm}
\indent In Fig. \ref{FER_SKR}, the FER (left) and the SKR (middle) are shown for different $\beta$ and $d$. For the FER we can see that using PS-256QAM for the modulation of the quantum states significantly impacts the error correction performance. Using the standard multi-dimensional protocol with $d = 8$, we lose 3.7\% in reconciliation efficiency when compared to an ideal BI-AWGN channel, which is equivalent to when $d = \infty$, for a FER of 10\%. Using higher dimensional reconciliation with $d=128$ allows us to close to gap by 2.6\%, with a gap of 1.1\% to the ideal case. The SKR we obtained when using $d = 128$ is 122\% higher when compared to $d = 8$, reducing the gap to BI-AWGN to 28\%. Although using this higher dimensional reconciliation is more complex, simplified versions of this protocol exist \cite{Jouguet_2011}, meaning that the bottleneck of the reconciliation is still the decoding on Alice's side.
\\\indent In Fig. \ref{FER_SKR} (right) we also show the SKR for our $\beta$-optimisation method and compare it to using the same $\beta$ for each CV-QKD block for FSO channels with different amounts of turbulence. The maximum SKR when using the same $\beta$ for each CV-QKD block is achieved at a $\beta$ around 93\% for all turbulence settings. As expected, the SKR is highest for the case where there is almost no turbulence. Between the other turbulence settings the SKR does not change significantly because of random fluctuations of $\xi_{Bob}$ during our measurements. When using our $\beta$ optimisation method (SKRs shown with dashed line), we can get up to 8.1\% higher SKR when compared to sticking to only one value for $\beta$. This gain is dependent on the specifications of the CV-QKD system, but considering that $\beta$-optimisation adds no additional complexity, while always increasing the SKR, it is always worth implementing.
\vspace{-6mm}
\section{Conclusion}
\vspace{-2mm}
\indent In this work we have demonstrated adaptive reconciliation for an experimental CV-QKD transmission over an FSO channel. We have shown that when using PS-256QAM for modulating the quantum states, it is worth considering using higher dimensional reconciliation to improve the SKR by up to 122\%. Finally, we proposed a method for optimising the reconciliation efficiency during parameter estimation, increasing the SKR by up to 8.1\%. 
\vspace{2mm}

\scriptsize \noindent 
This work was supported by the Dutch Ministry of Economic Affairs and Climate Policy (EZK), as part of the Quantum  Delta NL KAT-2 programme on Quantum Communications and PhotonDelta GrowthFunds Programme on Photonics.

 \vspace{-2mm}
\bibliographystyle{osajnl}
\bibliography{References} 

\end{document}